\documentclass[sigconf,authorversion]{acmart}

\usepackage{bm}
\usepackage{multirow}
\usepackage{mathtools}
\usepackage{subcaption}
\usepackage{paralist}
\usepackage[linesnumbered, ruled, vlined]{algorithm2e}
\SetKwInOut{Input}{Input}
\SetKwInOut{Output}{Output}

\newtheorem{problem}{Problem}

\newcommand{\onerewire}{{\ensuremath{1}}-{\sc Rewiring}\xspace}
\newcommand{\krewire}{{\ensuremath{k}}-{\sc Rewiring}\xspace}
\newcommand{\vertexcover}{{\sc Vertex\-Cover}\xspace}
\newcommand{\spara}[1]{\smallskip\noindent{\bf #1}}

\renewcommand{\vec}[1]{\mathbf{#1}}

\copyrightyear{2022}
\acmYear{2022}
\setcopyright{acmlicensed}
\acmConference[WWW '22] {Proceedings of the ACM Web Conference 2022}{April 25--29, 2022}{Virtual Event, Lyon, France.}
\acmBooktitle{Proceedings of the ACM Web Conference 2022 (WWW '22), April 25--29, 2022, Virtual Event, Lyon, France}
\acmPrice{15.00}
\acmISBN{978-1-4503-9096-5/22/04}
\acmDOI{10.1145/3485447.3512143}

\begin{document}

\title{Rewiring \emph{What-to-Watch-Next} Recommendations to Reduce Radicalization Pathways}

\author{Francesco Fabbri}
\affiliation{%
  \institution{Universitat Pompeu Fabra \& Eurecat}
  \city{Barcelona}
  \country{Spain}
}
\email{francesco.fabbri@eurecat.org}

\author{Yanhao Wang}
\orcid{0000-0002-7661-3917}
\affiliation{%
  \institution{East China Normal University}
  \city{Shanghai}
  \country{China}
}
\email{yhwang@dase.ecnu.edu.cn}

\author{Francesco Bonchi}
\orcid{0000-0001-9464-8315}
\affiliation{%
  \institution{ISI Foundation \city{Turin} \country{Italy}}
}
\affiliation{%
  \institution{Eurecat \city{Barcelona} \country{Spain}}
}
\email{francesco.bonchi@isi.it}

\author{Carlos Castillo}
\affiliation{%
  \institution{ICREA \& Universitat Pompeu Fabra}
  \city{Barcelona}
  \country{Spain}
}
\email{chato@icrea.cat}

\author{Michael Mathioudakis}
\orcid{0000-0003-0074-3966}
\affiliation{
  \institution{University of Helsinki}
  \city{Helsinki}
  \country{Finland}
}
\email{michael.mathioudakis@helsinki.fi}

\renewcommand{\shortauthors}{Fabbri et al.}

\begin{abstract}
Recommender systems typically suggest to users content similar to what they consumed in the past.
If a user happens to be exposed to strongly polarized content,  she might subsequently receive recommendations which may steer her towards more and more radicalized content, eventually being trapped in what we call a ``\emph{radicalization pathway}''.
In this paper, we study the problem of mitigating radicalization pathways using a graph-based approach.
Specifically, we model the set of recommendations of a ``\emph{what-to-watch-next}'' recommender as a $d$-regular directed graph where nodes correspond to content items, links to recommendations, and paths to possible user sessions.

We measure the ``\emph{segregation}'' score of a node representing radicalized content as the expected length of a random walk from that node to any node representing non-radicalized content.
High segregation scores are associated to larger chances to get users trapped in radicalization pathways.
Hence, we define the problem of reducing the prevalence of radicalization pathways by selecting a small number of edges to ``\emph{rewire}'', so to minimize the maximum of segregation scores among all radicalized nodes, while maintaining the relevance of the recommendations.

We prove that the problem of finding the optimal set of recommendations to rewire is NP-hard and NP-hard to approximate within any factor.
Therefore, we turn our attention to heuristics, and propose an efficient yet effective greedy algorithm based on the absorbing random walk theory.
Our experiments on real-world datasets in the context of video and news recommendations confirm the effectiveness of our proposal.
\end{abstract}



\maketitle

\section{Introduction}
\label{sec:intro}

``\emph{What-to-watch-next}'' (W2W) recommenders are key features of video sharing platforms~\cite{zhao2019recommending}, as they sustain user engagement, thus increasing content views and driving advertisement and monetization.
However, recent studies have raised serious concerns about the potential role played by W2W recommenders, specifically in driving users towards undesired or polarizing content~\cite{ledwich2019algorithmic}.
Specifically, radicalized communities\footnote{From~\citet{mccauley2008mechanisms}: ``\emph{Functionally, political radicalization is increased preparation for and commitment to intergroup conflict. Descriptively, radicalization means change in beliefs, feelings, and behaviors in directions that increasingly justify intergroup violence and demand sacrifice in defense of the ingroup.}''} on social networks and content sharing platforms have been recognized as keys in the consumption of news and in building opinions around politics and related subjects~\cite{rad_pat,roose2019making,weiss2018meet}.
Recent work highlights the role of recommender systems, which may steer users towards radicalized content, eventually building ``\emph{radicalization pathways}''~\cite{rad_pat,Ribeiro2020a} (i.e., a user might be further driven towards radicalized content even when this was not her initial intent).
In this paper, we study the problem of reducing the prevalence of radicalization pathways in W2W recommenders while maintaining the relevance of recommendations.

Formally, we model a W2W recommender system as a directed labeled graph where nodes correspond to videos (or other types of content) and directed edges represent recommendation links from one node to another\footnote{For ease of presentation, we focus on video sharing platforms. We note that the same type of recommendations occurs in many other contexts such as, for instance, news feeding platforms as shown in our experiments (see Section~\ref{sec:exp}).}.
In this scenario, each video is accompanied by the same number $d$ of recommendation links, and thus every node in the graph has the same out-degree $d$.
Moreover, each node has a binary label such as ``harmful'' (e.g., radicalized) or ``neutral'' (e.g., non-radicalized).
The browsing activity of a user through the W2W recommendations is modeled as a \emph{random walk} on the graph: after visiting a node (e.g., watching a video), the user moves to one of the $d$ recommended videos with a probability that depends on its visibility or ranking in the recommendation list.
In this setting, for each harmful node $v$, we measure the expected number of consecutive harmful nodes visited in a random walk before reaching any neutral node.
We call this measure the ``\emph{segregation}'' score of node $v$: intuitively, it quantifies how easy it is to get ``stuck'' in radicalization pathways starting from a given node.
Our goal is to reduce the segregation of the graph while guaranteeing that the quality of recommendations is maintained, where the quality is measured by the \emph{normalized discount cumulative gain}~\cite{DBLP:conf/sigir/BiegaGW18,DBLP:journals/tois/JarvelinK02} (nDCG) of each node.
An important challenge is that the underlying recommendation graph has intrinsically some level of homophily because, given that the W2W seeks to recommend relevant videos, it is likely to link harmful nodes to other harmful nodes.

We formulate the problem of reducing the segregation of the graph as selecting $k$ rewiring operations on edges (corresponding to modifications in the lists of recommended videos for some nodes) so as to minimize the maximum of segregation scores among all harmful nodes, while maintaining recommendation quality measured by nDCG above a given threshold for all nodes.
We prove that our \krewire problem is NP-hard and NP-hard to approximate within any factor.
We therefore turn our attention to design efficient and effective heuristics.
Our proposed algorithm is based on the \emph{absorbing random walk theory}~\cite{mavroforakis2015absorbing}, thanks to which we can efficiently compute the segregation score of each node and update it after every rewiring operation.
Specifically, our method finds a set of $k$ rewiring operations by greedily choosing the optimal rewiring for the special case of $k=1$ -- i.e., the \onerewire problem, then updates the segregation score of each node.
We further design a sorting and pruning strategy to avoid unnecessary attempts and thus improve the efficiency for searching the optimal rewiring.
Though the worst-case time complexity of our algorithm is quadratic with respect to the number of nodes $n$, it exhibits much better performance (nearly linear w.r.t.~$n$) in practice.

Finally, we present experiments on two real-world datasets: one in the context of video sharing and the other in the context of news feeds.
We compare our proposed algorithm against several baselines, including an algorithm for suggesting new edges to reduce radicalization in Web graphs.
The results show that our algorithm outperforms existing solutions in mitigating radicalization pathways in recommendation graphs.

In the rest of this paper, we first review the literature relevant to our work in Section~\ref{sec:rw}. Then, we introduce the background and formally define our problem in Section~\ref{sec:definition}. Our proposed algorithms are presented in Section~\ref{sec:alg}. The experimental setup and results are shown in Section~\ref{sec:exp}. Finally, we conclude this paper and discuss possible future directions in Section~\ref{sec:conclusion}.

\section{Related Work}
\label{sec:rw}

A great deal of research has been recently published about the potential created by unprecedented opportunities to access information on the Web and social media.
These risks include the spread of misinformation~\cite{10.1257/jep.31.2.211,DBLP:journals/sigkdd/ShuSWTL17}, the presence of bots ~\cite{DBLP:journals/cacm/FerraraVDMF16}, the abundance of offensive hate speech~\cite{DBLP:conf/ranlp/MalmasiZ17,DBLP:conf/ht/MondalSB17}, the availability of inappropriate videos targeting children~\cite{DBLP:conf/icwsm/PapadamouPZBKLS20}, the increase in controversy~\cite{DBLP:conf/wsdm/GarimellaMGM16} and polarization~\cite{DBLP:conf/icwsm/GuerraMCK13}, and the creation of radicalization pathways~\cite{Ribeiro2020a}.
Consequently, a substantial research effort has been devoted to model, detect, quantify, reduce, and/or block such negative phenomena.
Due to space limitations, we only discuss the existing studies that are the most relevant to our work here -- in particular, algorithmic approaches to optimizing graph structures for achieving the aforementioned goals~\cite{DBLP:conf/wsdm/GarimellaMGM17,DBLP:conf/www/MuscoMT18,DBLP:conf/wsdm/ChitraM20,10.1145/3437963.3441825,DBLP:conf/cikm/TongPEFF12,https://doi.org/10.1111/itor.12854,DBLP:conf/sdm/SahaAPV15,DBLP:conf/sdm/LeET15,DBLP:conf/kdd/ChenLB18,DBLP:journals/tkdd/YanLWLW19,DBLP:conf/kdd/KhalilDS14,DBLP:conf/icdm/KuhlmanTSMR13,DBLP:conf/aaai/KimuraSM08}.

A line of research deals with limiting the spread of undesirable content in a social network via edge manipulation~\cite{DBLP:conf/aaai/KimuraSM08,DBLP:conf/cikm/TongPEFF12,DBLP:conf/icdm/KuhlmanTSMR13,DBLP:conf/kdd/KhalilDS14,DBLP:conf/sdm/SahaAPV15,DBLP:conf/sdm/LeET15,DBLP:journals/tkdd/YanLWLW19}.
In these studies, the graph being manipulated is a network of users where the edges represent connections such as friendship or interactions among users.
In contrast, we consider a graph of content items (e.g., videos or news), where the edges represent recommendation links.
Moreover, these algorithmic methods are primarily based on information propagation models, while our work is based on random walks.

Another line of work aims at reducing controversy, disagreement, and polarization by edge manipulation in a social network, exposing users to others with different views~\cite{DBLP:conf/wsdm/GarimellaMGM17,DBLP:conf/kdd/ChenLB18,DBLP:conf/www/MuscoMT18,DBLP:conf/wsdm/ChitraM20,10.1145/3437963.3441825,https://doi.org/10.1111/itor.12854}.
\citet{DBLP:conf/wsdm/GarimellaMGM17} introduce the \emph{controversy score} of a graph based on random walks and propose an efficient algorithm to minimize it by edge addition.
\citet{DBLP:conf/www/MuscoMT18} introduce the \emph{Polarization-Disagreement index} of a graph based on Friedkin-Johnsen dynamics and propose a network-design approach to find a set of ``best'' edges that minimize this index.
\citet{DBLP:conf/kdd/ChenLB18} define the \emph{worst-case conflict risk} and \emph{average-case conflict risk} of a graph, also based on Friedkin-Johnsen dynamics, and propose algorithms to locally edit the graphs for reducing both measures.
\citet{DBLP:conf/wsdm/ChitraM20} analyze the impact of ``filter bubbles'' in social network polarization and how to mitigate them by graph modification.
\citet{https://doi.org/10.1111/itor.12854} define a polarization reduction problem by adding edges between users from different groups and propose integer programming-based methods to solve it.
Another related line of work proposes to model and mitigate the disparate exposure generated by people recommenders (e.g. who-to-follow link predictions) in presence of effects like homophily and polarization~\cite{fabbri2021exposure,fabbri2020effect,cinus2021effect,pitoura2020fairness}.
These studies also deal with networks of users, while in our case we consider a network of items.

The work probably most related to ours is the one by \citet{10.1145/3437963.3441825}, which considers a graph of items (e.g., Web pages with hyperlinks) and defines the structural bias of a node as the difficulty/effort needed to reach nodes of a different opinion.
They, then propose an efficient approximation algorithm to reduce the structural bias by edge insertions.
There are three main differences between this and our work.
First, two-directional edge manipulations (from class A to B and also from B to A) are considered by \citet{10.1145/3437963.3441825}, but one-directional edge manipulations (from harmful to neutral nodes only) are considered in our work.
Second, they consider inserting new links on a node, which better fits the case of Web pages, but we consider rewiring existing edges, which better fits the case of W2W recommenders.
Third, they define the structural bias of the graph as the sum of the bubble radii of all nodes, while we define the segregation of the graph as the worst-case segregation score among all harmful nodes.
We compare our proposed algorithm with theirs in our experiments.

A recent line of work introduces the notion of \emph{reachability} in recommender systems~\cite{dean2020recommendations,curmei2021quantifying}. Instead of rewiring the links, they focus on making allowable modifications in the user's rating history to avoid unintended consequences such as filter bubbles and radicalization.
However, as the problem formulation is different from ours, their proposed methods are not applicable to our problem.

Finally, there are many studies on modifying various graph characteristics, such as shortest paths~\cite{DBLP:conf/cikm/PapagelisBG11,DBLP:conf/sdm/ParotsidisPT15}, centrality~\cite{DBLP:conf/wsdm/ParotsidisPT16,DBLP:journals/tkdd/CrescenziDSV16,DBLP:conf/sdm/MedyaSSBS18,DBLP:journals/jea/BergaminiCDMSV18,DBLP:conf/aaai/DAngeloOS19,DBLP:conf/atal/WasWRM20}, opinion dynamics~\cite{DBLP:conf/kdd/AmelkinS19,DBLP:conf/aaai/CastiglioniF020}, and so on~\cite{DBLP:conf/sdm/ChanAT14,DBLP:journals/tkdd/Papagelis15,DBLP:journals/kais/LiY15,DBLP:conf/cikm/Zhu0WL18}, by edge manipulation.
We can draw insights from these methods but cannot directly apply them to our problem.

\section{Preliminaries}
\label{sec:definition}

Let us consider a set $V$ of $n$ items and a matrix $\vec S \in \mathbb{R}^{n \times n}$, where each entry $s_{uv} \in [0,1]$ at position $(u, v)$ denotes the relevance score of an item $v$ given that a user has browsed an item $u$.
This expresses the likelihood that a user who has just watched $u$ would be interested in watching $v$. 
Typically, a recommender system selects the $d$ most relevant items to compose the recommendation list $\Gamma^{+}(u)$ of $u$, where the number of recommendations $d$ is a design constraint (e.g., given by the size of the app window).
We assume that the system selects the top-$d$ items  $v$ w.r.t. $s_{uv}$ and that their relevance score uniquely determines the ranking of the $d$ items in $\Gamma^{+}(u)$. For each $v \in \Gamma^{+}(u)$, we use $i_{u}(v)$ to denote its ranking in $\Gamma^{+}(u)$.
After a user has seen $u$, she/he will find the next item to see from $\Gamma^{+}(u)$, and the probability $p_{uv}$ of selecting $v \in \Gamma^{+}(u)$ depends on the ranking $i_{u}(v)$ of $v$ in $\Gamma^{+}(u)$.
More formally, $p_{uv} = f(i_u(v))$, where $f$ is a non-increasing function that maps from $i_u(v)$ to $p_{uv}$ with $\sum_{v \in \Gamma^{+}(u)} p_{uv} = 1$.

This setting can be modeled as a directed probabilistic $d$-regular graph $G=(V,E, \vec M)$, where the node set $V$ corresponds to the set of all $n$ items, the edge set $E$ comprises $n \cdot d$ edges where each node $u \in V$ has $d$ out-edges connected to the nodes in $\Gamma^{+}(u)$, and $\vec M$ is an $n \times n$ transition matrix with a value of $p_{uv}$ for each $(u, v) \in E$ and $0$ otherwise. A user's browsing session is thus modeled as a random walk on $G$ starting from an arbitrary node in $V$ with transition probability $p_{uv}$ for each $(u, v) \in E$.

We further consider that the items in $V$ are divided into two disjoint subsets  $V_n$ and $V_h$ (i.e., $V_n \cap V_h = \emptyset$ and $V_n \cup V_h = V$) corresponding to ``neutral'' (e.g., not-radicalized) and ``harmful'' (e.g., radicalized) nodes, respectively.

The risk we want to mitigate is having users stuck in a long sequence of harmful nodes while performing a random walk.
In order to quantify this phenomenon we define the measure of \emph{segregation}.
Given a set $S \subset V$ of nodes and a node $u \in V \setminus S$, we use a random variable $T_u(S)$ to indicate the first instant when a random walk starting from $u$ reaches (or ``hits'') any node in $S$.
We define $\mathbb{E}_G[T_u(S)]$ as the \emph{hitting length} of $u$ w.r.t.~$S$, where the expectation is over the space of all possible random walks on $G$ starting from $u$. In our case, we define the segregation score $z_u$ of node $u \in V_h$ by its expected hitting length $\mathbb{E}_G[T_u(V_n)]$ w.r.t.~$V_n$.
The segregation $Z(G)$ of graph $G$ is defined by the maximum of segregation scores among all nodes in $V_h$ -- i.e., $Z(G) = \max_{u \in V_h} z_u$.
In the following, we omit the argument $G$ from $Z(G)$ when it is clear from the context.

Our main problem in this paper is to mitigate the effect of segregation by modifying the structure of $G$.
Specifically, we aim to find a set $O$ of rewiring operations on $G$, each of which removes an existing edge $(u, v) \in E$ and inserts a new one $(u, w) \notin E$ instead, such that $Z(G^O)$ is minimized, where $G^O$ is the new graph after performing $O$ on $G$.
For simplicity, we require that $u,v \in V_h$, $w \in V_n$, and $p_{uv}=p_{uw}$.
In other words, each rewiring operation changes the recommendation list $\Gamma^{+}(u)$ of $u$ by replacing one (harmful) item $v \in \Gamma^{+}(u)$ with another (neutral) item $w \notin \Gamma^{+}(u)$ and keeping the ranking $i_u(w)$ of $w$ the same as the ranking $i_u(v)$ of $v$ in $\Gamma^{+}(u)$.

Another goal, which is often conflicting, is to preserve the relevance of recommendations after performing the rewiring operations.
Besides requiring only a predefined number $k$ of rewirings, we also consider an additional constraint on the loss in the quality of the recommendations.
For this purpose we adopt the well-known \emph{normalized discounted cumulative gain} (nDCG)~\cite{DBLP:journals/tois/JarvelinK02,DBLP:conf/sigir/BiegaGW18} to evaluate the loss in the quality.
Formally, the \emph{discounted cumulative gain} (DCG) of a recommendation list $\Gamma^{+}(u)$ is defined as:
\begin{displaymath}
  \mathtt{DCG}(\Gamma^{+}(u)) = \sum_{v \in \Gamma^{+}(u)} \frac{s_{uv}}{1 + \log_{2}(1+i_u(v))}
\end{displaymath}
Then, we define the quality loss of $\Gamma^{+}(u)$ after rewiring operations by nDCG as follows:
\begin{equation}\label{eq:dcg}
  L(\Gamma^{+}(u)) = \mathtt{nDCG}(\Gamma^{+}(u)) = \frac{\mathtt{DCG}(\Gamma^{+}(u))}{\mathtt{DCG}(\Gamma^{+}_{0}(u))}
\end{equation}
where $ \Gamma^{+}_{0}(u) $ is the original (ideal) recommendation list where all the top-$d$ items that are the most relevant to $u$ are included.

Let $o=(u, v, w)$ be a rewiring operation that deletes $(u, v)$ while adding $(u, w)$ and $O$ be a set of rewiring operations. For ease of presentation, we define a function $ \Delta(O) \triangleq Z(G) - Z(G^O) $ to denote the decrease in the segregation after performing the rewiring operations in $O$ and updating $G$ to $G^O$. We are now ready to formally define the main problem studied in this paper.
\begin{problem}[\krewire]
  Given a directed probabilistic graph $G=(V,E,\vec M)$, a positive integer $k \in \mathbb{Z}^+$, and a threshold $\tau \in (0,1)$, find a set $O$ of $k$ rewiring operations that maximizes $\Delta(O)$, under the constraint that $L(\Gamma^{+}(u)) \geq \tau$ for each node $u \in V$.
  \label{problem:krewire}
\end{problem}

The hardness of the \krewire problem is analyzed in the following theorem.
\begin{theorem}\label{thm:np:hard}
  The \krewire problem is NP-hard and NP-hard to approximate within any factor.
\end{theorem}
We show the NP-hardness of the \krewire problem by reducing from the \vertexcover problem. Furthermore, we show that finding an $\alpha$-approximate solution of the \krewire problem for any factor $\alpha > 0$ is at least as hard as finding the minimum vertex cover of a graph. Therefore, the \krewire problem is NP-hard to approximate within any factor.
The proof of Theorem~\ref{thm:np:hard} can be found in Appendix~\ref{proof:np:hard}.

\subsection{Absorbing Random Walk}
\label{subsec:arw}

We now provide notions from the absorbing random walk theory~\cite{mavroforakis2015absorbing} on which our algorithms are built.

The \krewire problem asks to minimize \emph{segregation}, which is defined as the maximum hitting length from any harmful node to neutral nodes. %
Specifically, in the context of \krewire for the given probabilistic directed graph $G=(V,E,\vec M)$, we equivalently consider a modified transition matrix $\vec M$ as follows:
\begin{equation*}
  \vec M =
  \begin{bmatrix}
    \vec M_{hh} & \vec M_{hn} \\
    \vec 0 & \vec I
  \end{bmatrix}
\end{equation*}
In the matrix $\vec M$ above, each \emph{neutral} node has been set to be \emph{absorbing}, i.e., its transition probability to itself is set to $p_{ii} = 1$ and $0$ to other nodes (see the bottom row of $\vec M$). Intuitively, no random walk passing through an absorbing node can move away from it~\cite{mavroforakis2015absorbing}.
For each \emph{harmful} node, its transition probabilities remain unmodified (see the top row of $\vec M$) and thus the node remains \emph{transient} (i.e., \emph{non-absorbing}).

The fundamental matrix $\vec F$ can be computed from the sub-matrix $\vec M_{hh}$ as follows~\cite{mavroforakis2015absorbing}:
\begin{equation*}
  \vec F = (\vec I - \vec M_{hh})^{-1}
\end{equation*}
where the entry $f_{uv}$ represents the expected total number of times that the random walk visits node $v$ having started from node $u$. Then, the expected length of a random walk that starts from any node and stops when it gets absorbed is given by vector $\vec z$:
\begin{equation}\label{eq:z}
  \vec z =
  \begin{bmatrix}
    \vec (\vec I - \vec M_{hh})^{-1} \\
    \vec 0
  \end{bmatrix}
  \vec 1
\end{equation}
where $\vec 1$ is an $n$-dimensional vector of all $1$'s.
Here, the $i$-th entry $z_i$ of vector $\vec z$ represents the expected number of random walk steps before being absorbed by any absorbing node, assuming that the random walk starts from the $i$-th node.

Given that the absorbing and transient nodes are set to correspond exactly to the neutral and harmful nodes, respectively, the values of $\vec z$ correspond exactly to the expected hitting length as used to define segregation. 
Hence, the \krewire problem asks to choose a set of $k$ rewiring operations to minimize the maximum entry $Z = \max_{1 \leq i \leq n} z_i$ of vector $\vec z$.

\section{Algorithms}
\label{sec:alg}

Since \krewire is NP-hard to approximate within any factor, we propose an efficient heuristic.
The heuristic is motivated by the following observation: despite the NP-hardness of \krewire, its special case when $k=1$, which we call \onerewire, is solvable in polynomial time.
Given an optimal \onerewire algorithm, \krewire can be addressed by running it $k$ times.

We begin our presentation of algorithms by showing a brute-force algorithm for finding the optimal solution of \onerewire (Section~\ref{subsec:brute-force}), as well as a way to speed it up via incremental updates (Section~\ref{subsec:incremental}).
Subsequently, we propose our optimal \onerewire algorithm that improves the efficiency of the brute-force algorithm by faster rewiring search (Section~\ref{subsec:onerewire}).
Finally, we present how our \onerewire algorithm is used for \krewire (Section~\ref{subsec:krewires}).

\subsection{Brute-Force Algorithm for \texorpdfstring{\onerewire}{1-Rewiring}}
\label{subsec:brute-force}

Given a graph $G$ and a rewiring operation $o$, we use $\Delta(o)$ to denote the decrease in $Z$ after performing $o$ on $G$.
We present a brute-force algorithm to find the rewiring operation $o^*$ that maximizes $\Delta(o)$.
The algorithm has three steps:
\begin{inparaenum}[(1)]
\item enumerate the set $\Omega$ of all feasible rewiring operations for $G$ and a given threshold $\tau$;
\item get $\Delta(o)$ for each $o \in \Omega$ by computing $Z$ using Eq.~\ref{eq:z} on $G$ before/after performing $o$;
\item find the operation $o$ that has the largest $\Delta(o)$ as the optimal solution $o^*$.
\end{inparaenum}
In the brute-force algorithm, since the number of existing edges is $O(d n)$ and the number of possible new edges to rewire is $O(n)$ for each existing edge, the size of $\Omega$ is $O(d n^2)$.
In addition, the old and new values of $Z$ can be computed by matrix inversion using Eq.~\ref{eq:z} in $O(n^3)$ time. Therefore, the brute-force algorithm runs in $O(d n^5)$ time.
As all feasible operations are examined, this solution is guaranteed to be optimal.

The brute-force algorithm is impractical if the graph is large, due to the huge number of feasible operations and the high cost of computing $Z$.
We introduce two strategies to improve its efficiency.
First, we update the vector $\vec z$ incrementally for a rewiring operation.
Second, we devise efficient strategies to avoid unnecessary computation when searching for the optimal rewiring operation, leading to our optimal \onerewire algorithm.

\subsection{Incremental Update of Vector \texorpdfstring{$\vec z$}{z}}
\label{subsec:incremental}

We analyze how the fundamental matrix $\vec F$ and vector $\vec z$ change after performing a rewiring operation $o=(u, v, w)$.
Two edits will be performed on $G$ for $o$:
\begin{inparaenum}[(1)]
\item the removal of an existing edge $(u, v) \in E$ and
\item the insertion of a new edge $(u, w) \notin E$ to $E$.
\end{inparaenum}

The two operations update the transition matrix $\vec M$ to $\vec M'$ as follows:
\begin{displaymath}
  \vec M' = \vec M + \vec e \vec g^{\top}
\end{displaymath}
where $\vec e$ is an $n$-dimensional column vector that indicates the position of the source node $u$: %
\begin{displaymath}
  \begin{aligned}
    e_{j}= \left\{
      \begin{array}{ll}
        1 & \text{if} \; j=u \\
        0 & \text{otherwise}
      \end{array}
    \right.
  \end{aligned}
\end{displaymath}
and $\vec g^{\top}$ is an $n$-dimensional row vector that denotes the changes in the transition probabilities.
Specifically, for the removal of $(u, v)$ and insertion of $(u, w)$, the probability $p_{uv}$ of $(u, v)$ is reassigned
to $(u, w)$. We denote the probability as $p_o = p_{uv}$. Formally,
\begin{displaymath}
  \begin{aligned}
    g_{j}= \left\{
      \begin{array}{ll}
        - p_o & \text{if} \; j=v \\
        + p_o & \text{if} \; j=w \\
        0 & \text{otherwise}
      \end{array}
    \right.
  \end{aligned}
\end{displaymath}
Thus, operation $o=(u, v, w)$ on the fundamental matrix $\vec F$ yields an updated fundamental matrix $\vec F'$:
\begin{displaymath}
  \vec F' = ((\vec I - \vec M_{hh}) - \vec e \vec g^{\top})^{-1} = (\vec F^{-1} + (-1) \vec e \vec g^{\top})^{-1}
\end{displaymath}
By applying the Sherman-Morrison formula~\cite{press2007numerical}, we can avoid the computation of the new inverse
and express $\vec F'$ as:
\begin{equation}\label{eq:diff:F}
  \vec F' = \vec F - \frac{\vec F \vec e \vec g^{\top} \vec F}{1 + \vec g^{\top} \vec F \vec e}
\end{equation}
Accordingly, the new vector $\vec z'$ is expressed as:
\begin{equation}\label{eq:diff}
  \vec z' = \vec z - \frac{\vec F \vec e \vec g^{\top} \vec F}{1 + \vec g^{\top} \vec F \vec e} \vec 1
\end{equation}
The denominator of the second term in Eq.~\ref{eq:diff} can be written as:
\begin{displaymath}
  1 + \vec g^{\top} \vec F \vec e = 1 - p_{o}(f_{wu} \cdot \bm{1}_{w \in V_h} - f_{vu})
\end{displaymath}
where $\bm{1}_{w \in V_h}$ is an indicator that is equal to $1$ if $w \in V_h$ and $0$ otherwise.
Because, as mentioned in Section~\ref{sec:definition}, we restrict ourselves to rewiring with $w \not\in V_h$, the above expression is simplified as:
\begin{displaymath}
  1 + \vec g^{\top} \vec F \vec e = 1 + p_{o}f_{vu}.
\end{displaymath}
Meanwhile, the numerator of the second term in Eq.~\ref{eq:diff} is written as:
\begin{displaymath}
  \vec F \vec e \vec g^{\top} \vec F \vec 1 = -\vec f_u (z_w \cdot \bm{1}_{w \in V_h} - z_v) p_{o}
\end{displaymath}
where $\vec f_u$ is the column vector corresponding to $u$ in $\vec F$,
$z_{w}$ and $z_{v}$ are the entries of $\vec z$ for $u$ and $v$, respectively.
As previously, because $w \not\in V_h$, we have that Eq.~\ref{eq:diff} is simplified as:
\begin{displaymath}
  \vec z'  = \vec z - \frac{\vec f_{u} z_v}{1/p_{o} + f_{vu}}.
\end{displaymath}
For any harmful node $h$, we calculate its decrease $\Delta(h, o)$ in segregation score
after performing $o = (u, v, w)$ as:
\begin{equation}\label{eq:delta:z}
 \Delta(h, o) = z_h - z^{\prime}_{h} = \frac{f_{hu} z_v}{1/p_{o} + f_{vu}}
\end{equation}
The optimal \onerewire we present next is based on Eq.~\ref{eq:delta:z}.

\subsection{Optimal \texorpdfstring{\onerewire}{1-Rewiring} Algorithm}
\label{subsec:onerewire}

\begin{algorithm}[t]
  \small
  \caption{\textsc{Optimal} \onerewire}
  \label{algo:one-rewiring}
  \Input{Graph $G = (V, E, \vec M)$, fundamental matrix $\vec F$, segregation vector $\vec z$, threshold $\tau$}
  \Output{Optimal rewiring operation $o^{*}$}
  Initialize $\Omega \gets \emptyset$, $o^* \gets NULL$, $\Delta^* \gets 0$\;
  \ForEach{node $u \in V_h$\label{ln:cand:s}}{
    Find node $w \in V_n$ s.t.~$(u, w) \notin E$ and $s_{uw}$ is the maximum\;
    \ForEach{node $v \in V_h$ with $(u, v) \in E$}{
      Add $o=(u, v, w)$ to $\Omega$ if $L(\Gamma^{+}(u)) \geq \tau$ after replacing $(u, v)$ with $(u, w)$\;
    }
    \label{ln:cand:t}
  }
  Sort nodes in $V_h$ as $\langle h_1, \ldots, h_{n_h} \rangle$ in descending order of $z_h$\;\label{ln:search:s}
  \ForEach{$o \in \Omega$}{
    Compute $\Delta(h_1, o)$ using Eq.~\ref{eq:delta:z}\;
    \uIf{$z^{\prime}_{h_1} > z_{h_2}$}{
      $\Delta(o) \gets \Delta(h_1, o)$\;
    }
    \Else{
      Find the largest $j > 1$ such that $z^{\prime}_{h_1} < z_{h_j}$\;
      Compute $\Delta(h_i, o)$ for each $i=2,\ldots,j$\;
      $\Delta(o) \gets z_{h_1} - \max_{i \in [1,j]} z^{\prime}_{h_i}$\;
    }
    \If{$\Delta(o) > \Delta^*$}{
      $o^* \gets o$ and $\Delta^* \gets \Delta(o)$\;
    }
    \label{ln:search:t}
  }
  \Return{$o^{*}$}\;
\end{algorithm}

We now introduce our method to find the optimal solution $o^*$ of \onerewire, i.e., the rewiring operation
that maximizes $\Delta(o)$ among all $o \in \Omega$.
The detailed procedure is presented in Algorithm~\ref{algo:one-rewiring}, to which the fundamental matrix $\vec F$ and segregation vector $\vec z$ are given as input.
The algorithm proceeds in two steps:
\begin{inparaenum}[(1)]
\item candidate generation, as described in Lines~\ref{ln:cand:s}--\ref{ln:cand:t}, which returns a set $\Omega$ of possible rewiring operations that definitely include the optimal \onerewire, and
\item optimal rewiring search, as described in Lines~\ref{ln:search:s}--\ref{ln:search:t}, which computes the objective value for each candidate rewiring to identify the optimal one.
\end{inparaenum}
Compared with the brute-force algorithm, this method reduces the cost of computing $\Delta(o)$ since it only probes a few nodes with the largest segregation scores.
In addition, it can still be guaranteed to find the optimal solution,
as all rewiring operations that might be the optimal one have been considered.

\spara{Candidate generation.}
The purpose of this step is to exclude from enumeration all rewiring operations that violate the quality constraint. %
Towards this end, we do not consider any rewiring operation that for any node $u$ will lead to
the \emph{discount cumulative gain} (DCG) of $u$ below the threshold $\tau$.
According to Eq.~\ref{eq:delta:z}, we find that $\Delta(h, o)$ of node $h$ w.r.t.~$o=(u, v, w)$ is independent of $(u, w)$.
Therefore, for a specific node $u$, we can fix $w$ to
the neutral (absorbing) node with the highest relevance score $s_{uw}$ and $(u, w) \notin E$
so that as many rewiring operations as possible are feasible.
Then, we should select the node $v$ where $(u, v) \in E$ will be replaced.
We need to guarantee that $L(\Gamma^{+}(u)) \geq \tau$ after $(u, v)$ is replaced by $(u, w)$.
For each node $v \in \Gamma^{+}(u)$, we can take $s_{uv}$ and $s_{uw}$ into Eq.~\ref{eq:dcg}.
If $L(\Gamma^{+}(u)) \geq \tau$, we will list $o=(u, v, w)$ as a candidate.
After considering each node $u \in V_h$, we generate the set $\Omega$
of all candidate rewiring operations.

\spara{Optimal rewiring search.}
The second step is to search for the optimal rewiring operation $o^*$ from $\Omega$.
We first sort all harmful nodes in descending order of their segregation scores as $\langle h_1, h_2, \ldots, h_{n_h} \rangle$, where $h_i$ is the node with the $i$-th largest segregation score.
Since we are interested in minimizing the maximum segregation, we can focus on the first few nodes with the largest segregation scores and ignore the remaining ones.
We need to compute $\Delta(o)$ for each $o \in \Omega$ and always keep the maximum of $\Delta(o)$.
After evaluating every $o \in \Omega$, it is obvious that the one maximizing $\Delta(o)$ is $o^*$.
Furthermore, to compute $\Delta(o)$ for some operation $o$, we perform the following steps:
\begin{inparaenum}[(1)]
\item compute $\Delta(h_1, o)$ using Eq.~\ref{eq:delta:z};
\item if $z^{\prime}_{h_1} > z_{h_2}$, then $\Delta(o) = \Delta(h_1, o)$;
\item otherwise, find the largest $j$ such that $z^{\prime}_{h_1} < z_{h_j}$, compute $\Delta(h_i, o)$ for each $i=2,\ldots,j$; in this case,
we have $\Delta(o) = z_{h_1} - \max_{i \in [1,j]} z^{\prime}_{h_i}$.
\end{inparaenum}

\spara{Time complexity.}
Compared with the brute-force algorithm, the size of $\Omega$ is reduced from $O(d n^2)$ to $O(d n)$.
Then, sorting the nodes in $V_h$ takes $O(n \log{n})$ time.
Moreover, it takes $O(1)$ time to compute $\Delta(h, o)$ for each $h$ and $o$.
For each $o \in \Omega$, $\Delta(h, o)$ is computed $O(n)$ times in the worst case.
Therefore, the time complexity is $O(d n^2)$ in the worst case.
However, in our experimental evaluation, we find that $\Delta(h, o)$ is computed only a small number of times.
Therefore, if computing $\Delta(o)$ takes $O(1)$ time in practice, then the anticipated running time is $O \big( n (d + \log{n}) \big)$, as confirmed empirically.

\begin{algorithm}[t]
  \small
  \caption{\textsc{Heuristic} \krewire}
  \label{algo:heuristic}
  \Input{Graph $G = (V, E, \vec M)$, threshold $\tau$, size constraint $k$}
  \Output{A set $O$ of $k$ rewiring operations}
  Compute the initial $\vec F$ and $\vec z$ based on $\vec M$\;
  Acquire $\Omega$ using Lines~\ref{ln:cand:s}--\ref{ln:cand:t} of Algorithm~\ref{algo:one-rewiring}\;
  Initialize $O \gets \emptyset$\;
  \For{$i \gets 1,2,\ldots,k$}{
    Run Lines~\ref{ln:search:s}--\ref{ln:search:t} of Algorithm~\ref{algo:one-rewiring} to get $o^*=(u^*,v^*,w^*)$\;
    $O \gets O \cup \{o^*\}$\;
    Update $G$, $\vec M$, $\vec F$, and $\vec z$ for $o^*$\;
    Delete the existing rewiring operations of $u^*$ from $\Omega$
    and add new possible operations of $u^*$ to $\Omega$\;
    \If{$\Omega = \emptyset$}{
      \textbf{break}\;
    }
  }
  \Return{$O$}\;
\end{algorithm}

\subsection{Heuristic \texorpdfstring{\krewire}{k-Rewiring} Algorithm}
\label{subsec:krewires}

Our \krewire algorithm based on the \onerewire algorithm is presented in Algorithm~\ref{algo:heuristic}.
Its basic idea is to find the $k$ rewiring operations by running the \onerewire algorithm $k$ times.
The first step is to initialize the fundamental matrix $\vec F$ and segregation vector $\vec z$. In our implementation,
instead of performing the expensive matrix inversion (in Eq.~\ref{eq:z}), $\vec F$ and $\vec z$ are approximated
through the power iteration method in~\cite{mavroforakis2015absorbing}.
Then, the procedure of candidate generation is the same as that in Algorithm~\ref{algo:one-rewiring}.
Next, it runs $k$ iterations for getting $k$ rewiring operations.
At each iteration, it also searches for the the optimal rewiring operation $o^*=(u^*,v^*,w^*)$ among $\Omega$
as Algorithm~\ref{algo:one-rewiring}.
After that, $G$, $\vec M$, $\vec F$, and $\vec z$ are updated according to $o^*$
(see Eq.~\ref{eq:diff:F} and~\ref{eq:diff} for the update of $\vec F$ and $\vec z$).
Since the existing rewiring operations of $u^*$ are not feasible anymore, it will regenerate new possible operations of $u^*$ based on the updated $\Gamma^{+}(u^*)$ and the threshold $\tau$ to replace the old ones.
Finally, the algorithm terminates when $k$ rewiring operations have been found or there is no feasible operation anymore.

\spara{Time complexity.}
The time complexity of computing $\vec F$ and $\vec z$
is $O(\operatorname{iter} \cdot d n)$ where $\operatorname{iter}$ is the number of iterations in the power method.
The time to update $\vec F$ and $\vec z$ for each rewiring operation is $O(n)$.
Overall, its time complexity is $O(k d n^2)$ since it is safe to consider that $\operatorname{iter} \ll n$.
In practice, it takes $O(1)$ time to compute $\Delta(o)$ and $\operatorname{iter} = O(k)$,
and thus the running time of the \krewire algorithm can be regarded as $O \left(k n \left(d + \log{n}\right) \right)$.

\section{Experiments}
\label{sec:exp}

Our experiments aim to:
\begin{inparaenum}[(1)]
\item show the effectiveness of our algorithm on mitigating radicalization pathways compared to existing algorithms;
\item test the robustness of our algorithm with respect to different thresholds $\tau$; and
\item illustrate how much our algorithm can reduce the total exposure to harmful content.
\end{inparaenum}

\subsection{Experimental Setup}

\textbf{Datasets.}
We perform experiments within two application domains: video sharing and news feeding.

For the first application, we use the \textbf{YouTube} dataset~\cite{Ribeiro2020a}, which contains 330,925 videos and 2,474,044 recommendations.
The dataset includes node labels such as ``\emph{alt-right}'', ``\emph{alt-lite}'', ``\emph{intellectual dark web}'' and ``\emph{neutral}''.
We categorize the first three classes as ``radicalized'' or harmful and the last class as ``neutral,'' following the analysis done by this dataset's curators~\cite{Ribeiro2020a}, in which these three classes are shown to be overlapping in terms of audience and content.
When generating the recommendation graphs, we consider only videos having a minimum of 10k views.
In this way, we filter out all the ones with too few interactions.
We consider the video-to-video recommendations collected via simulations as implicit feedback interactions, where the video-to-video interactions can be formatted as a square matrix, with position $(u, v)$ containing the number of times the user jumped from video $u$ to video $v$.
Using alternating least squares (ALS)~\cite{hu2008collaborative}, we can first derive the latent dimensions of the matrix, generate the scores (normalized to $[0,1]$) and then build the recommendation lists for each video.
We eventually create different $d$-regular graphs with $d \in \{5, 10, 20\}$.
To evaluate the effect of graph size on performance, we also use a smaller subset of videos with only 100k or more views for graph construction.
Finally, we have 3 smaller (\textbf{YT-D5-S}, \textbf{YT-D10-S}, and \textbf{YT-D20-S}) and 3 larger (\textbf{YT-D5-B}, \textbf{YT-D10-B}, and \textbf{YT-D20-B}) recommendation graphs.

For the second application, we use the \textbf{NELA-GT} dataset~\cite{norregaard2019nela}, which is a collection of 713k news in English.
Each news article includes title, text, and timestamp, as well as credibility labels (reliable or unreliable).
Our task is to reduce the risk of users getting stuck in unreliable content via \emph{``what-to-read-next''} recommendations.
To build the recommendation graphs, we compute the pairwise semantic similarities between news through the pre-generated weights with RoBERTa~\cite{liu2019roberta}.
After normalizing the scores in the range $[0,1]$, in order to reproduce different instances of news feeding websites, we generate different subsets of news by month.
We perform our experiments on the 4 months with the largest number of news: August (\textbf{NEWS-1}), September (\textbf{NEWS-2}), October (\textbf{NEWS-3}) and November (\textbf{NEWS-4}).

\begin{table}[t]
\centering
\caption{Characteristics of the recommendation graphs used in the experiments, including out-degree $d$, number of nodes $n$, number of edges $m$, fraction of nodes from $V_{h}$ (i.e., $n_h/n$), and initial segregation $Z^0$ of each graph.}
\label{tab:summary}
\begin{tabular}{cccccc}
\toprule
\multicolumn{6}{c}{\textbf{YouTube}} \\
\midrule
\textbf{Name} & $d$ & $n$ & $m$ & $n_h/n$ & $Z^0$ \\
\midrule
YT-D5-S  & \multirow{2}{*}{5}  & 31524  & 157620  & 0.48 & 588.86 \\
YT-D5-B  &                     & 105143 & 525715  & 0.43 & 598.32 \\
\midrule
YT-D10-S & \multirow{2}{*}{10} & 31524  & 315240  & 0.48 & 718.92 \\
YT-D10-B &                     & 105143 & 1051430 & 0.43 & 718.37 \\
\midrule
YT-D20-S & \multirow{2}{*}{20} & 31524  & 630480  & 0.48 & 328.03 \\
YT-D20-B &                     & 105143 & 2102860 & 0.43 & 331.09 \\
\bottomrule
\toprule
\multicolumn{6}{c}{\textbf{NELA-GT}} \\
\midrule
\textbf{Name} & $d$ & $n$ & $m$ & $n_h/n$ & $Z^0$ \\
\midrule
NEWS-1   & \multirow{4}{*}{10} & 27286  & 272860  & 0.61 & 88.53 \\
NEWS-2   &                     & 22296  & 222960  & 0.62 & 29.90 \\
NEWS-3   &                     & 28861  & 288610  & 0.61 & 335.23 \\
NEWS-4   &                     & 26114  & 261140  & 0.65 & 75.15 \\
\bottomrule
\end{tabular}
\end{table}

The characteristics of the ten recommendation graphs used in our experiments are reported in Table~\ref{tab:summary}.

\newcommand{\rbl}{\emph{RePBubLik}\xspace}

\spara{Algorithms.}
We compare our proposed heuristic (\textbf{HEU}) algorithm for \krewire with three baselines and one existing algorithm.
The first baseline (\textbf{BSL-1}) selects the set of $k$ rewiring operations by running Algorithm 1. Instead of picking only one rewiring operation, it picks the $k$ operations with the largest values of $\Delta$ all at once.
The second baseline (\textbf{BSL-2}) considers the best possible $k$ rewiring operations by looking at the initial values of the vector $\vec z$. It firsts select the $k$ nodes with the largest $z$ values, then among the possible rewiring operations from those nodes, it returns the $k$ operations with the largest values of $\Delta$.
The third baseline (\textbf{RND}) just picks $k$ random rewiring operations from all the candidates.
Finally, the existing method we compare with is the \rbl algorithm~\cite{10.1145/3437963.3441825} (\textbf{RBL}). It reduces the structural bias of the graph by looking at the bubble radius of the two partitions of nodes, returning a list of $k$ new edges to add. The original algorithm is designed for the insertion of new links, and not for the rewiring (deletion + insertion).
Consequently, we adapt the \rbl algorithm to our objective as follows:
\begin{inparaenum}[(1)]
\item we run it to return a list of potential edges to be added for reducing the structural bias of the harmful nodes;
\item for each potential insertion, in order to generate a rewiring operation, we check among the existing edges to find the one edge that meets the quality constraint $\tau$ after being replaced by the new edge;
\item we finally select a set of $k$ rewiring operations from the previous step.
\end{inparaenum}

The experiments were conducted on a server running Ubuntu 16.04 with an Intel Broadwell 2.40GHz CPU and 29GB of memory.
Our algorithm and baselines were implemented in Python 3.
Our code and datasets are publicly available at \url{https://github.com/FraFabbri/rewiring-what-to-watch}.
The implementation of \rbl is available at \url{https://github.com/CriMenghini/RePBubLik}.

\subsection{Experimental Results}

\begin{figure}[t]
  \centering
  \captionsetup[sub]{skip=0pt}
  \subcaptionbox*{}[\linewidth]{\includegraphics[width=3in]{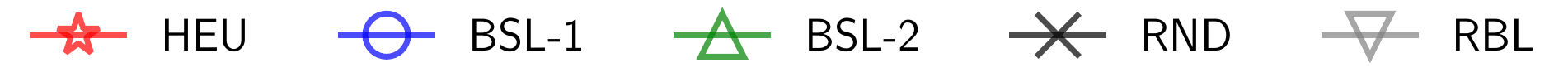}}
  \\
  \vspace{-1em}
  \subcaptionbox{YT-D5-B}[.475\linewidth]{\includegraphics[width=1.5in]{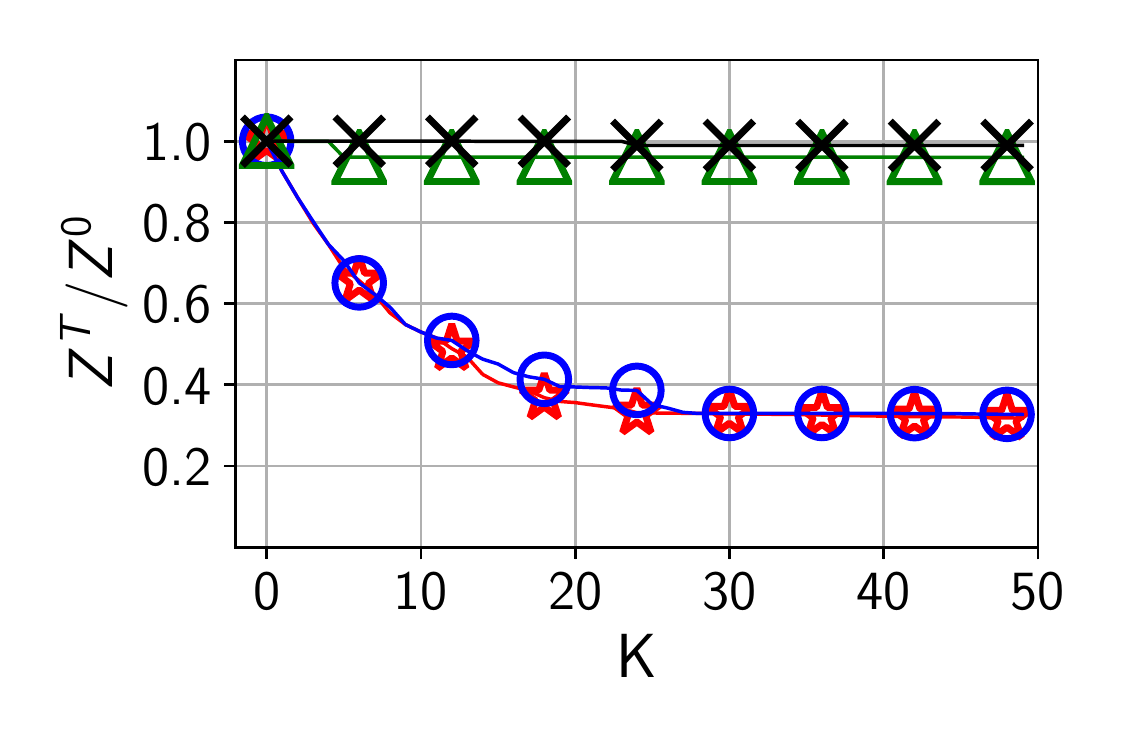}}
  \subcaptionbox{YT-D5-S}[.475\linewidth]{\includegraphics[width=1.5in]{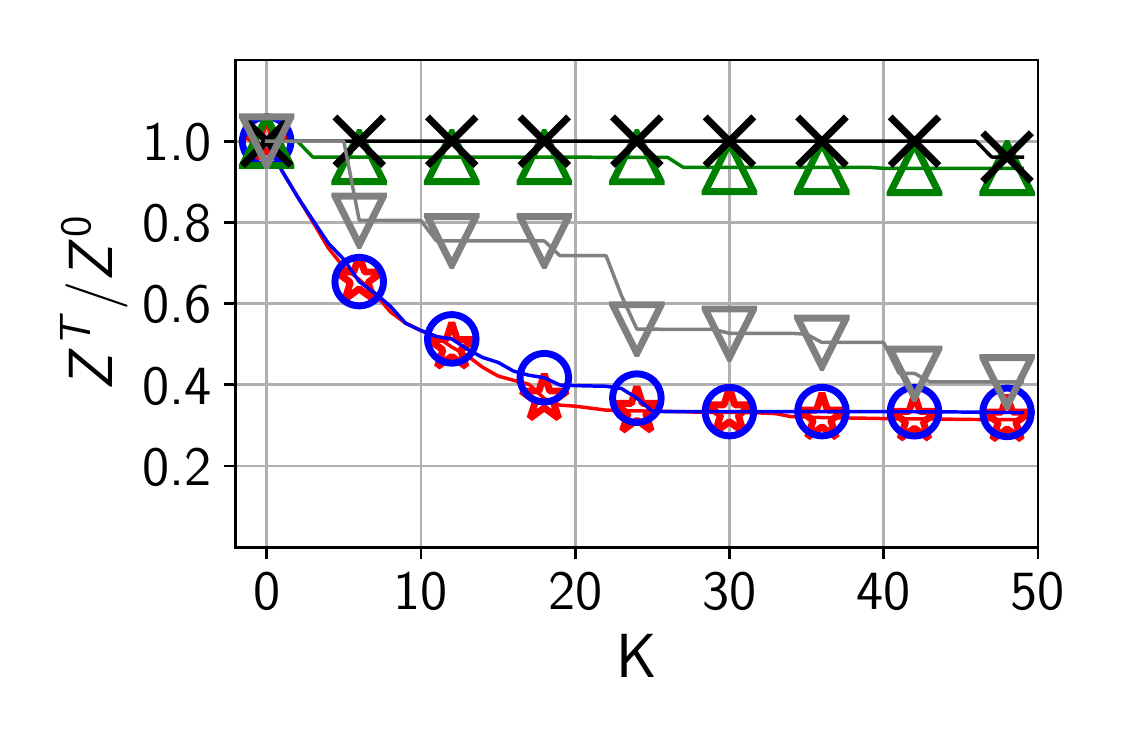}}
  \\
  \subcaptionbox{YT-D10-B}[.475\linewidth]{\includegraphics[width=1.5in]{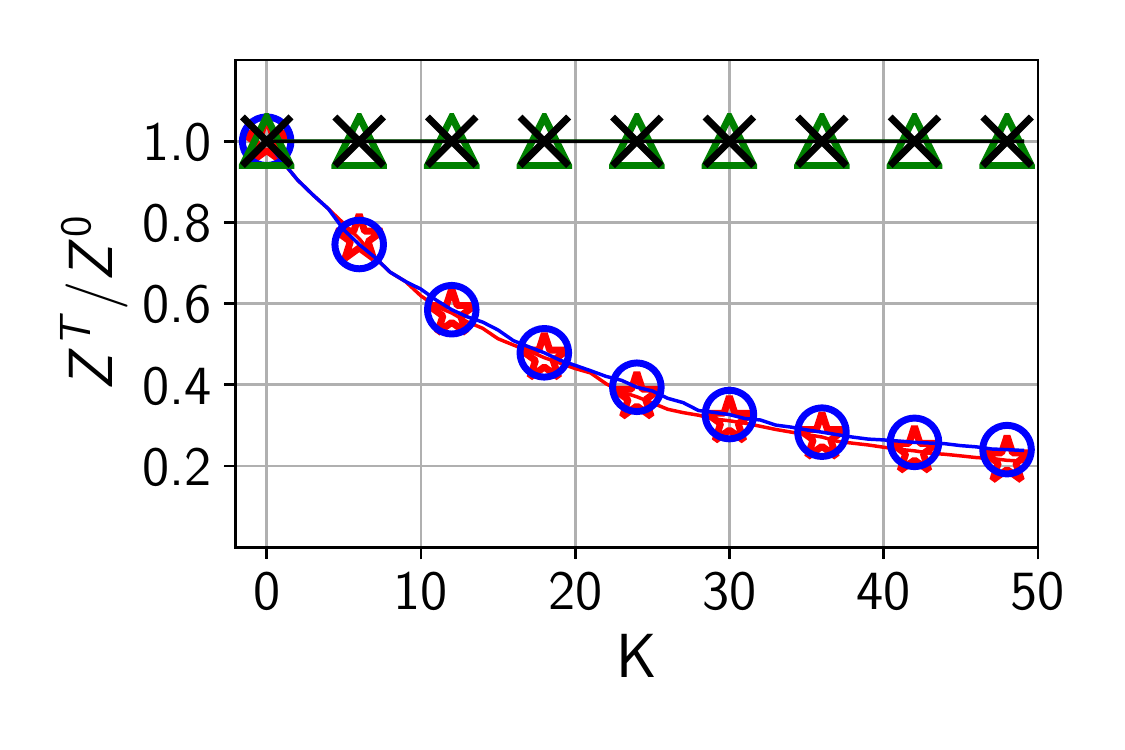}}
  \subcaptionbox{YT-D10-S}[.475\linewidth]{\includegraphics[width=1.5in]{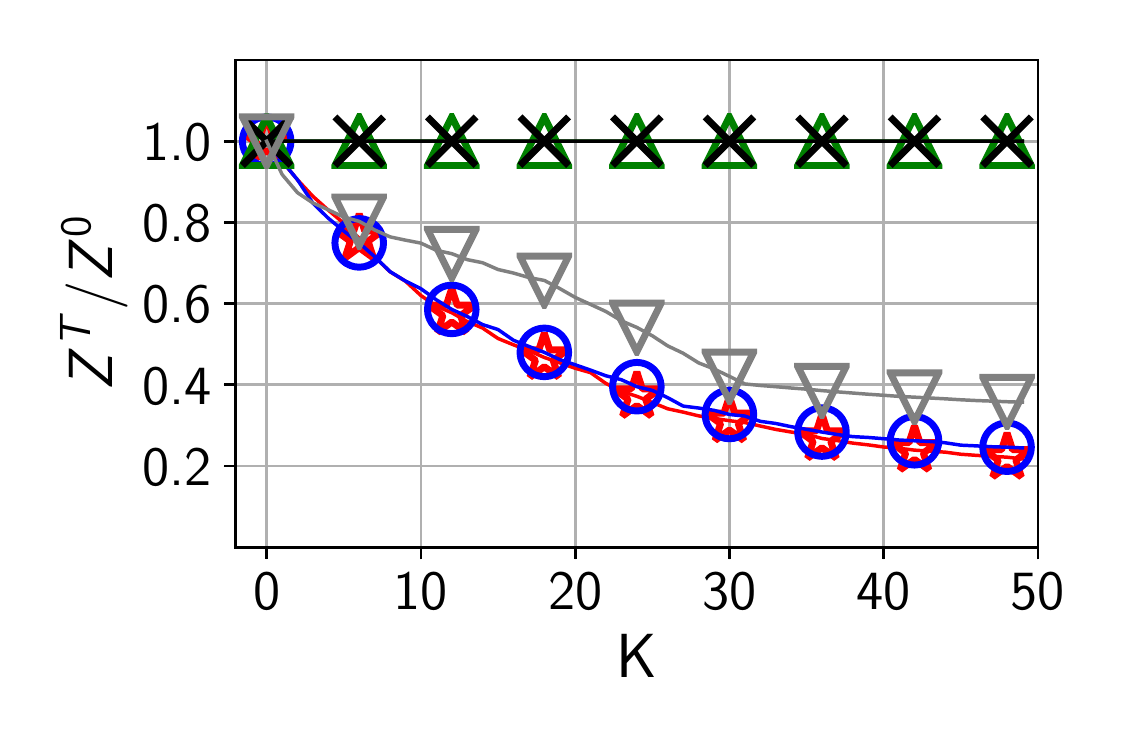}}
  \\
  \subcaptionbox{YT-D20-B}[.475\linewidth]{\includegraphics[width=1.5in]{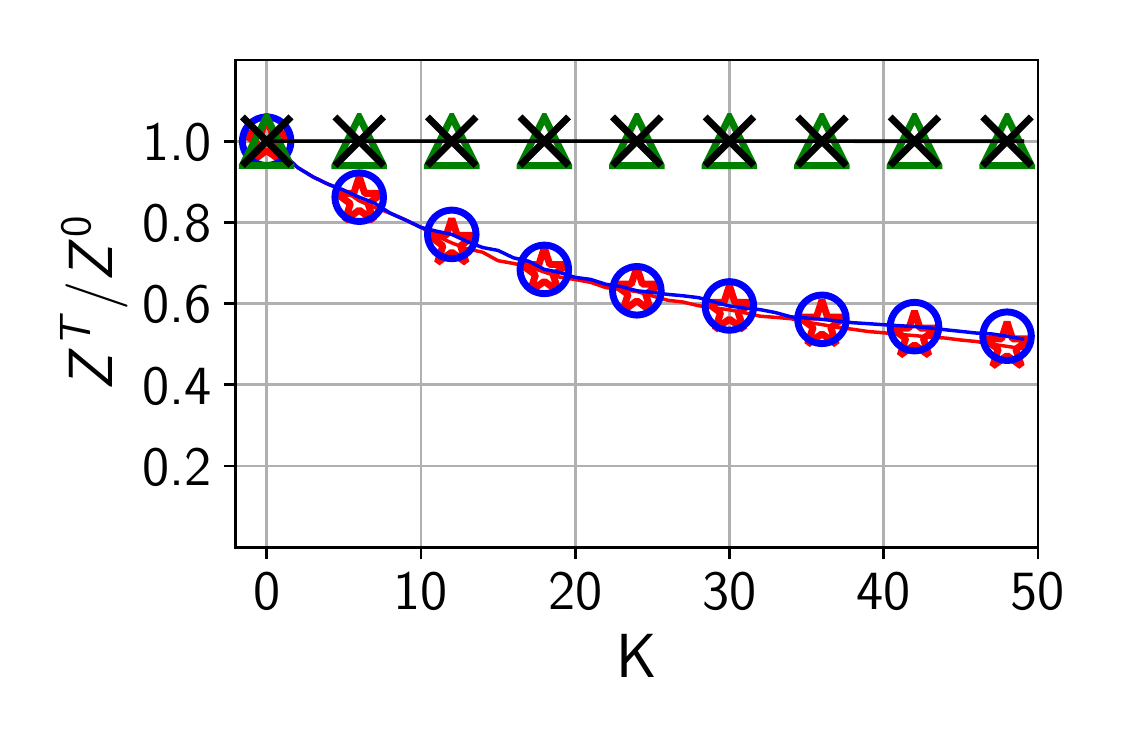}}
  \subcaptionbox{YT-D20-S}[.475\linewidth]{\includegraphics[width=1.5in]{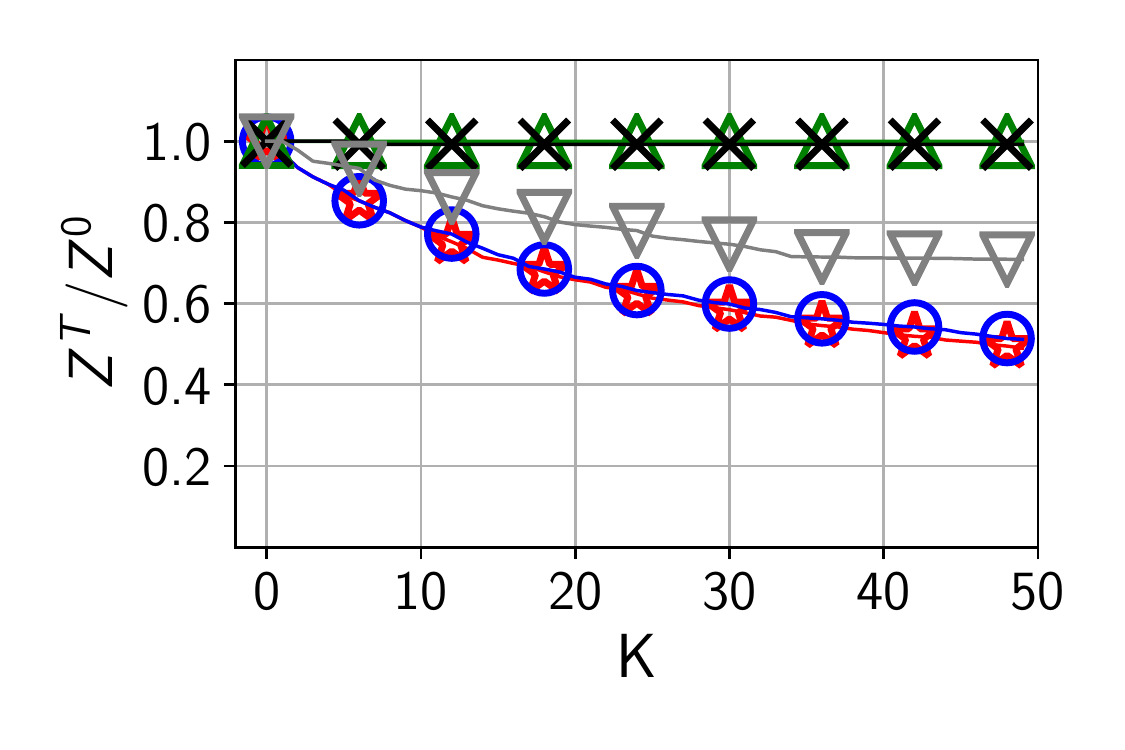}}
  \caption{Performance comparison in the YouTube dataset.}
  \Description{Figure 1}
  \label{fig:youtube-res}
\end{figure}

\textbf{Effectiveness of our method.}
In Figure~\ref{fig:youtube-res}, we present the results on the YouTube recommendation graphs.
On each graph, we evaluate the performance of each algorithm along 50 rewiring operations with the threshold of quality constraint is fixed to $\tau = 0.9$.
We keep track of the relative decrease in the segregation $Z^T/Z^0$ after each rewiring operation, where $Z^0$ is the initial segregation and $Z^T$ is the segregation after $T$ rewiring operations.
On all the graphs, it is clear that our heuristic algorithm (\textbf{HEU}) outperforms all the competitors.
On the graphs with the smallest out-degree ($d=5$), it decreases $Z$ by over $40\%$ within only 10 rewiring operations (i.e., $Z^{10}/Z^{0} \leq 0.6$).
In this case, it stops decreasing $Z$ after 30 rewiring operations, which implies that only after a few rewiring operations our heuristic algorithm has found the best possible operations constrained by the threshold $\tau$.
On the graphs with $d = 10$, our heuristic algorithm is able to decrease $Z$ by nearly $80\%$, which is even larger than the case of $d=5$.
This result is consistent in both smaller (YT-D10-S) and bigger (YT-D10-B) graphs. On the graphs with the largest out-degree ($d=20$), the algorithm is still effective but, as expected, achieves a comparable reduction in $Z$ after 50 operations.

The first baseline (\textbf{BSL-1}) shows almost the same solution quality as \textbf{HEU}, since most of the operations found by both algorithms are the same. Although the rewiring operations provided by \rbl (\textbf{RBL}) also decrease the original $Z_0$ significantly, they are less effective than the ones given by our algorithm.
Also, with a smaller size of recommendation list ($d=5$), it reaches some steady states along the iterations, where the new rewiring operations do not decrease the $Z$ value at all.
For the YouTube dataset, we present only the results of \textbf{RBL} on the smaller graphs (the second column of Figure~\ref{fig:youtube-res}), since it cannot finish in reasonable time (24 hours) on larger graphs. The other baseline (\textbf{BSL-2}) and the random solution (\textbf{RND}) do not produce substantial decreases over the initial $Z_0$.

\begin{figure}[t]
  \centering
  \captionsetup[sub]{skip=0pt}
  \subcaptionbox*{}[\linewidth]{\includegraphics[width=3in]{plots/youtube/fixed-tau/legend.pdf}}
  \\
  \vspace{-1em}
  \subcaptionbox{NEWS-1}[.475\linewidth]{\includegraphics[width=1.5in]{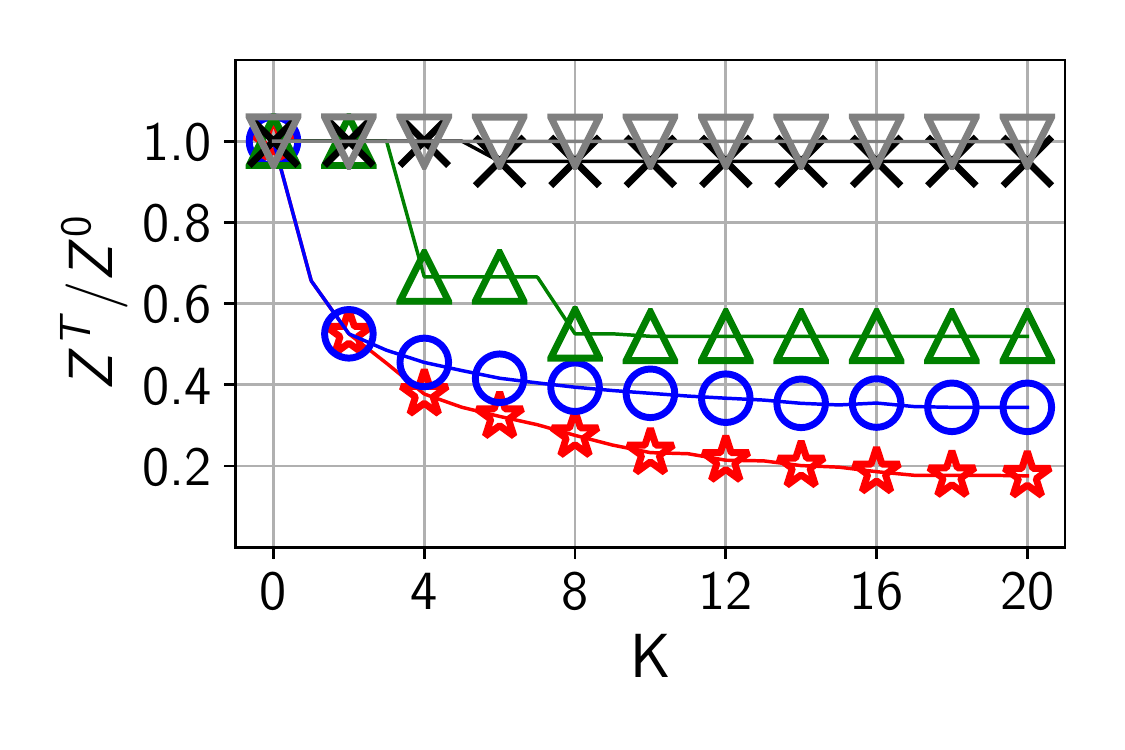}}
  \subcaptionbox{NEWS-2}[.475\linewidth]{\includegraphics[width=1.5in]{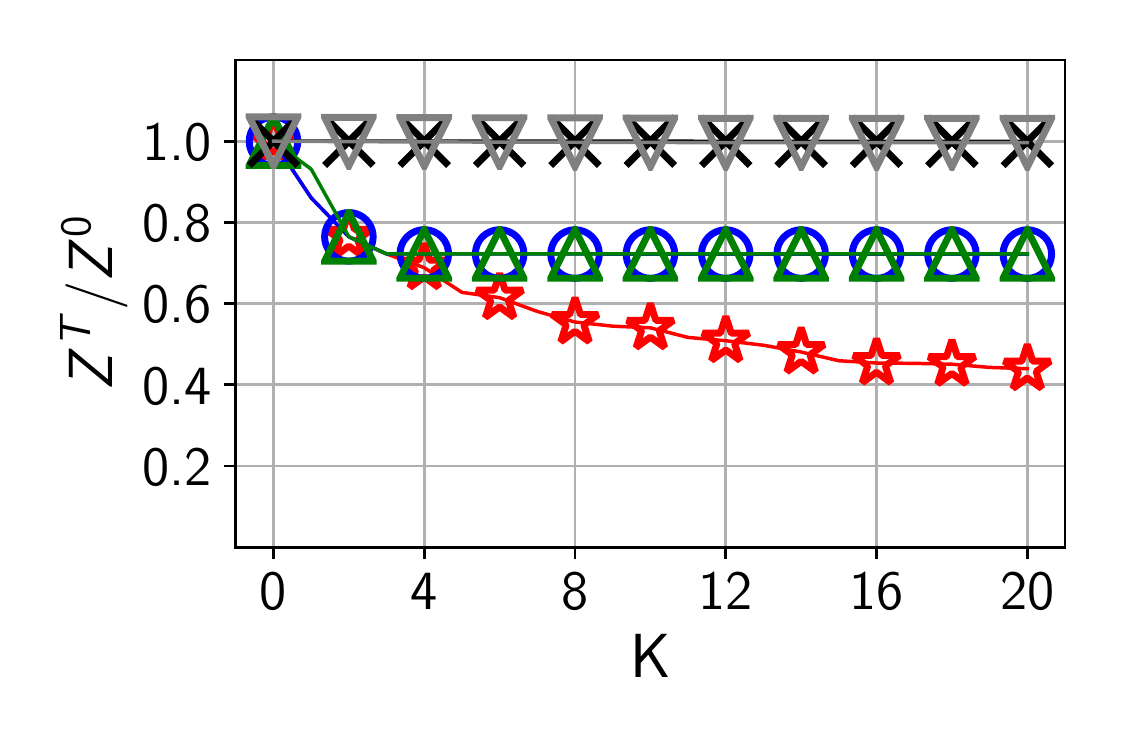}}
  \\
  \subcaptionbox{NEWS-3}[.475\linewidth]{\includegraphics[width=1.5in]{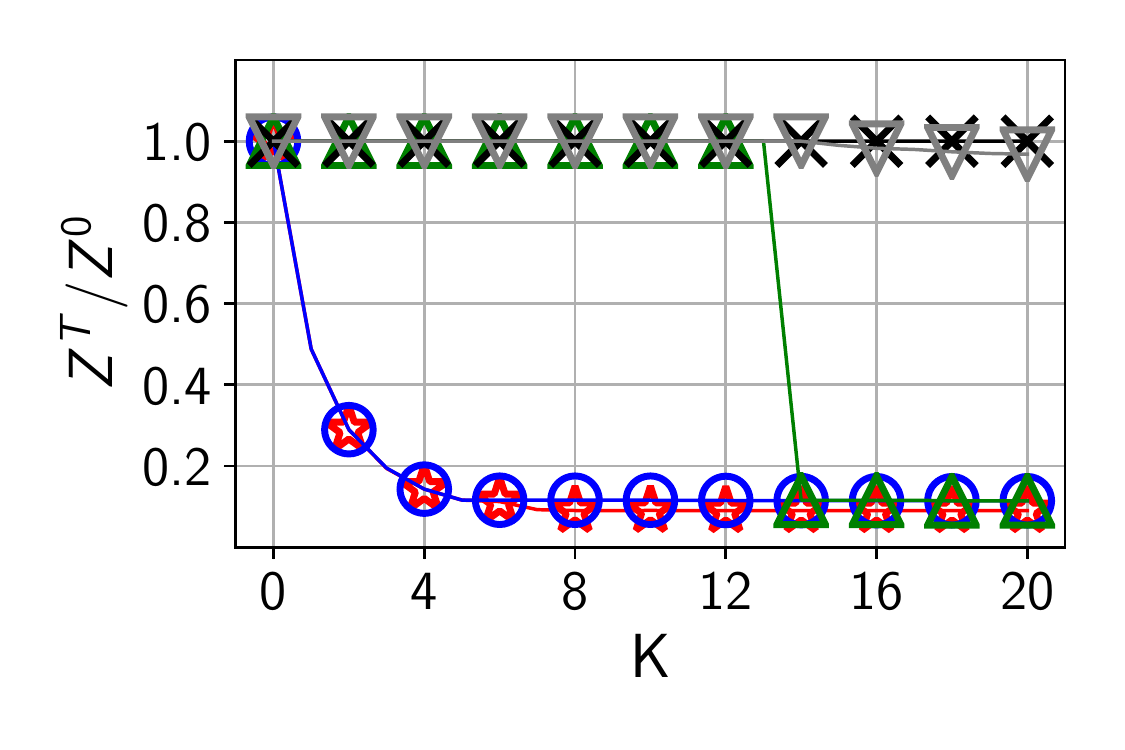}}
  \subcaptionbox{NEWS-4}[.475\linewidth]{\includegraphics[width=1.5in]{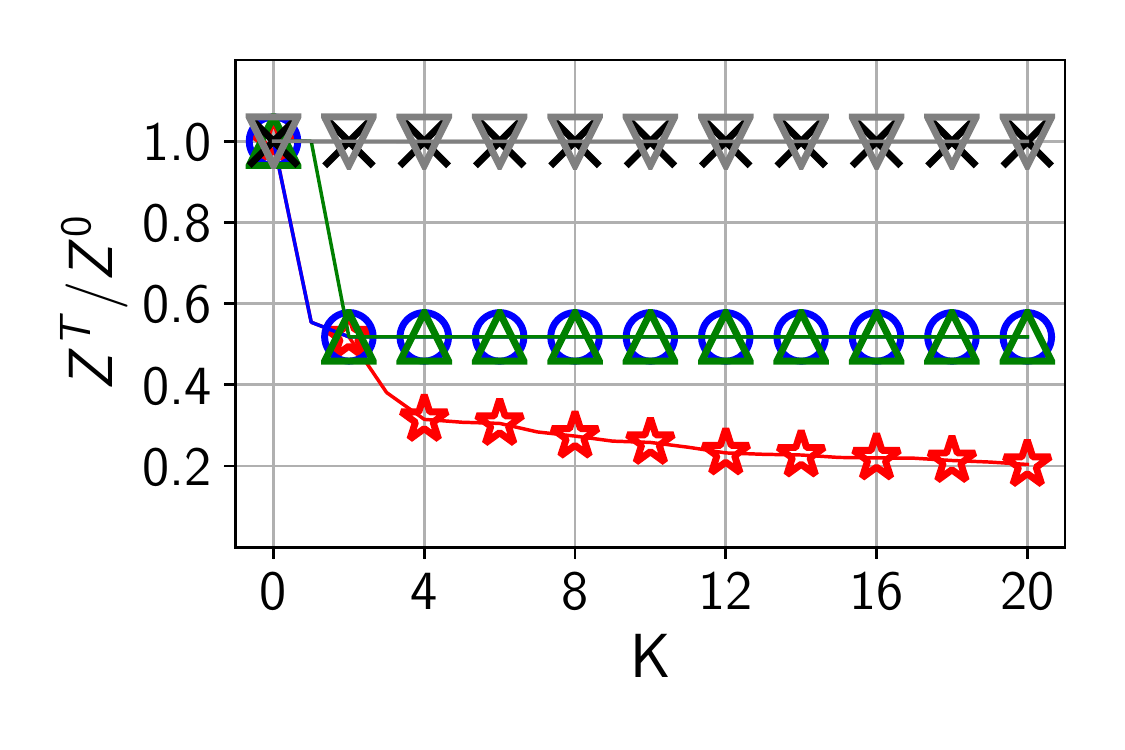}}
  \caption{Performance comparison in the NELA-GT dataset.}
  \Description{Figure 2}
  \label{fig:news-res}
\end{figure}

In Figure~\ref{fig:news-res}, we present the results on the NELA-GT recommendation graphs.
We also fix $\tau = 0.9$ in these experiments.
Given that the values of $Z^0$ are smaller in the news recommendation graph, we evaluate the performance of different algorithms with smaller $k$ (i.e., $k = 20$).
As for the previous case, our heuristic algorithm is the one achieving the best performance on every graph, which reduces $Z$ by at least $60\%$ after 20 rewiring operations.
Furthermore, on the graph with the biggest $Z$ value (NEWS-3), it decreases the initial segregation by more than $80\%$ only after 4 rewiring operations.
The two baselines (\textbf{BSL-1} and \textbf{BSL-2}) show comparable performance, but only on NEWS-3 they obtain close drops in $Z_0$ to \textbf{HEU} after 20 iterations. In the other cases, they are stuck in steady states far from \textbf{HEU}.
The rewiring provided by \rbl (\textbf{RBL}) shows no significant decrease over the initial $Z_0$, which is comparable only to \textbf{RND}.
The difference in performance between YouTube and NELA-GT can be to some extent attributed to differences in their degree distributions.
We compute the Gini coefficient of the in-degree distribution of the graphs: for the YouTube graphs the Gini coefficient of in-degree for the harmful nodes is never below $90\%$; while for the NELA-GT graphs this index is never above $50\%$.
These differences imply that \rbl might not perform well when the in-degree distribution of the graph is not highly skewed.

\begin{figure}[t]
  \centering
  \captionsetup[sub]{skip=0pt}
  \subcaptionbox*{}[\linewidth]{\includegraphics[width=3in]{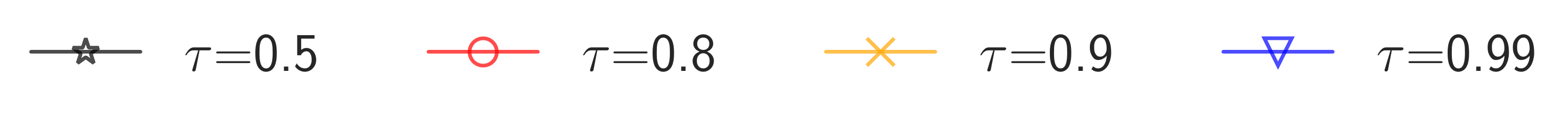}}
  \\
  \vspace{-1em}
  \subcaptionbox{YT-D5-B}[.475\linewidth]{\includegraphics[width=1.5in]{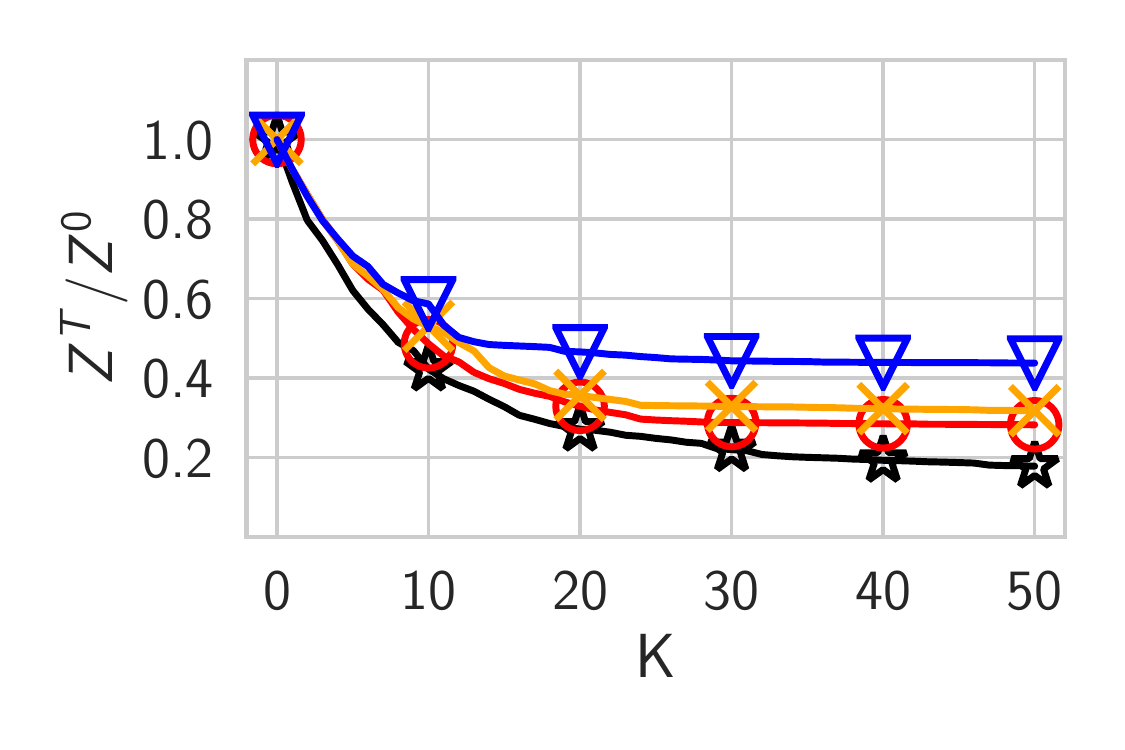}}
  \subcaptionbox{YT-D5-S}[.475\linewidth]{\includegraphics[width=1.5in]{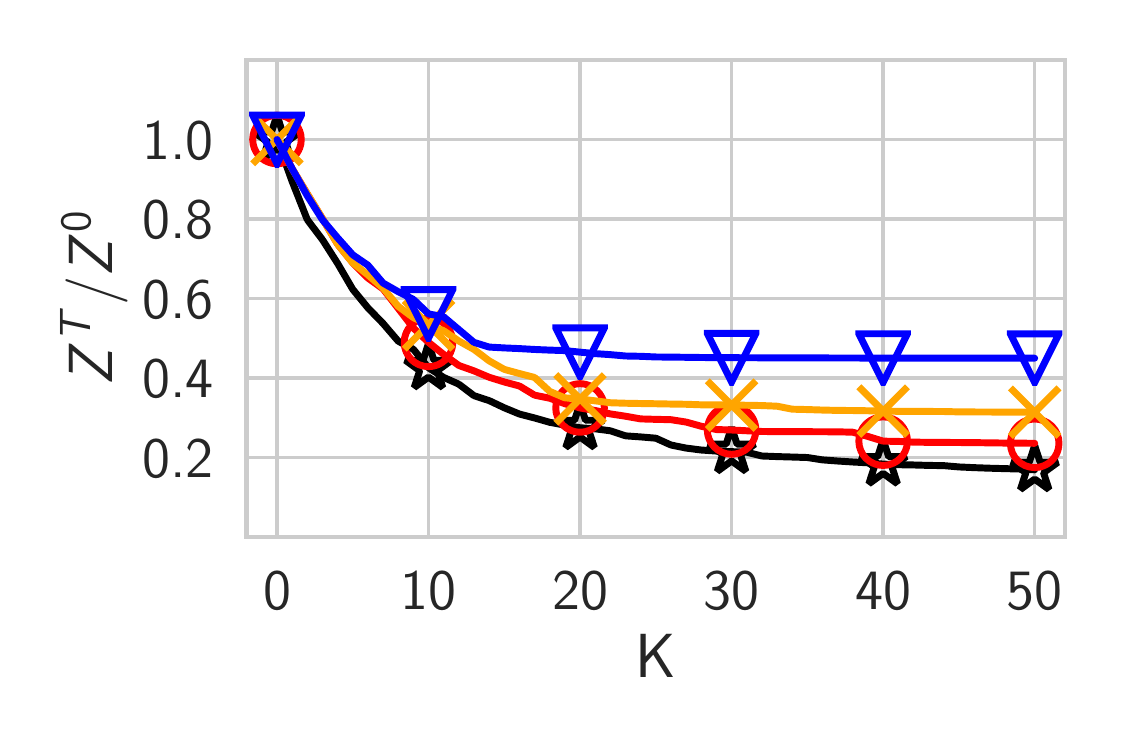}}
  \\
  \subcaptionbox{YT-D20-B}[.475\linewidth]{\includegraphics[width=1.5in]{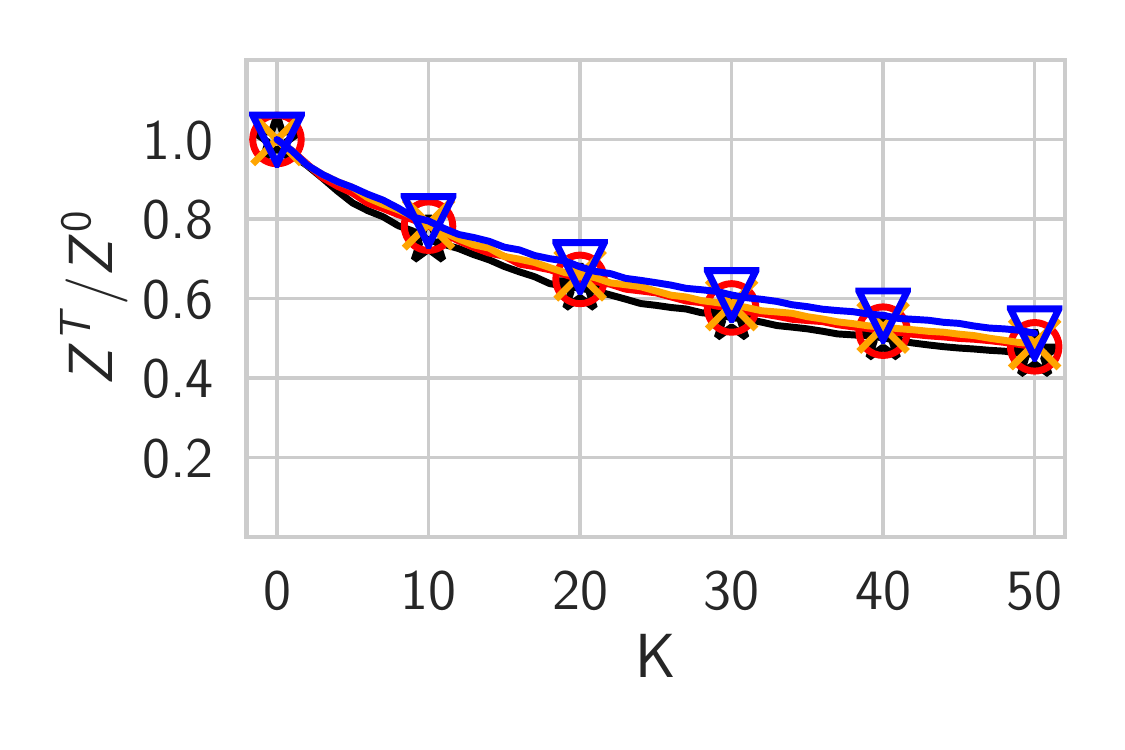}}
  \subcaptionbox{YT-D20-S}[.475\linewidth]{\includegraphics[width=1.5in]{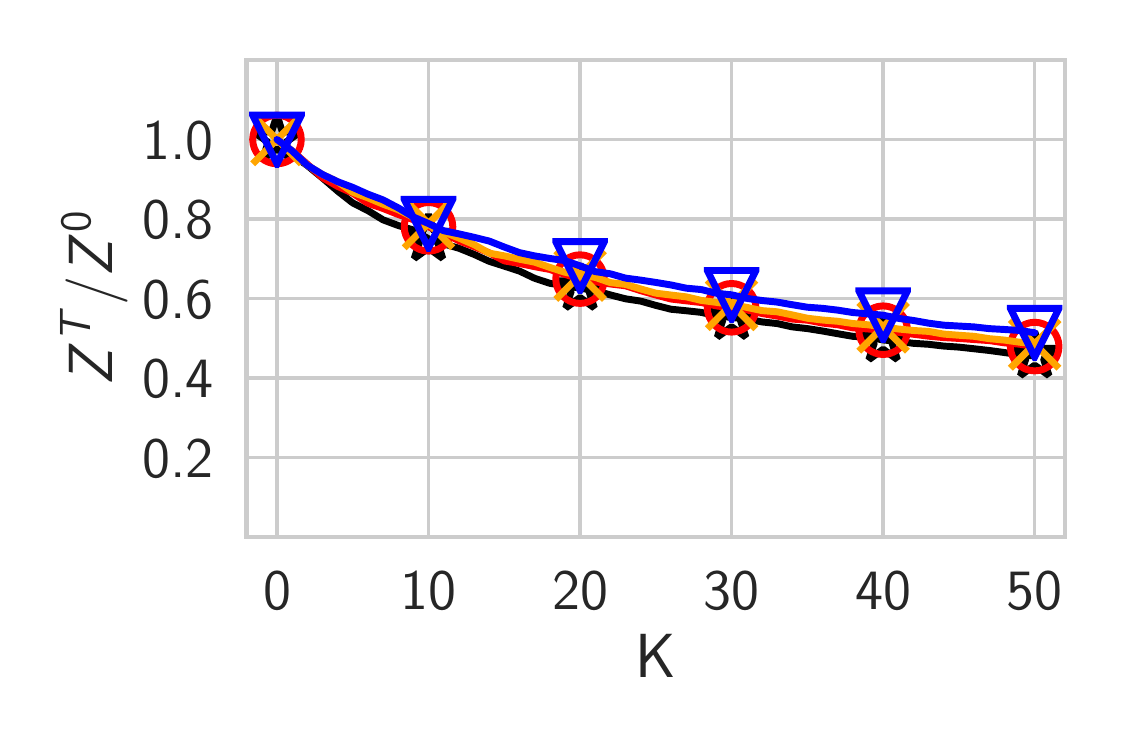}}
  \caption{Performance of our algorithm (HEU) with varying quality constraints $\tau$.}
  \Description{Figure 3}
  \label{fig:tau-res}
\end{figure}

\spara{Robustness w.r.t.~threshold of recommendation quality.}
To investigate the role of the threshold $\tau$ of recommendation quality on the output of our algorithm, we test on the YouTube recommendation graphs with the same number of rewiring operations ($k = 50$) but different values of $\tau$ in $\{0.5, 0.8, 0.9, 0.99\}$.
We present the results in Figure~\ref{fig:tau-res}.
As expected, under a more lenient quality constraint ($\tau = 0.5$), the algorithm achieves a larger decease in the value of $Z$.
It is also clear that the differences are less evident on graphs with a larger out-degree ($d=20$). Specifically, for a smaller out-degree ($d=5$) all the $\tau$ configurations except $\tau = 0.5$ tend to stabilize after $k=20$ rewiring operations. This is because the number of possible rewiring operations constrained by $\tau$ is small.
It is also evident that the graph size, given different values of $\tau$, does not impact the overall performance of our algorithm.

\begin{figure}[t]
  \centering
  \captionsetup[sub]{skip=0pt}
  \subcaptionbox{Heuristic (HEU)}[.475\linewidth]{\includegraphics[width=1.5in]{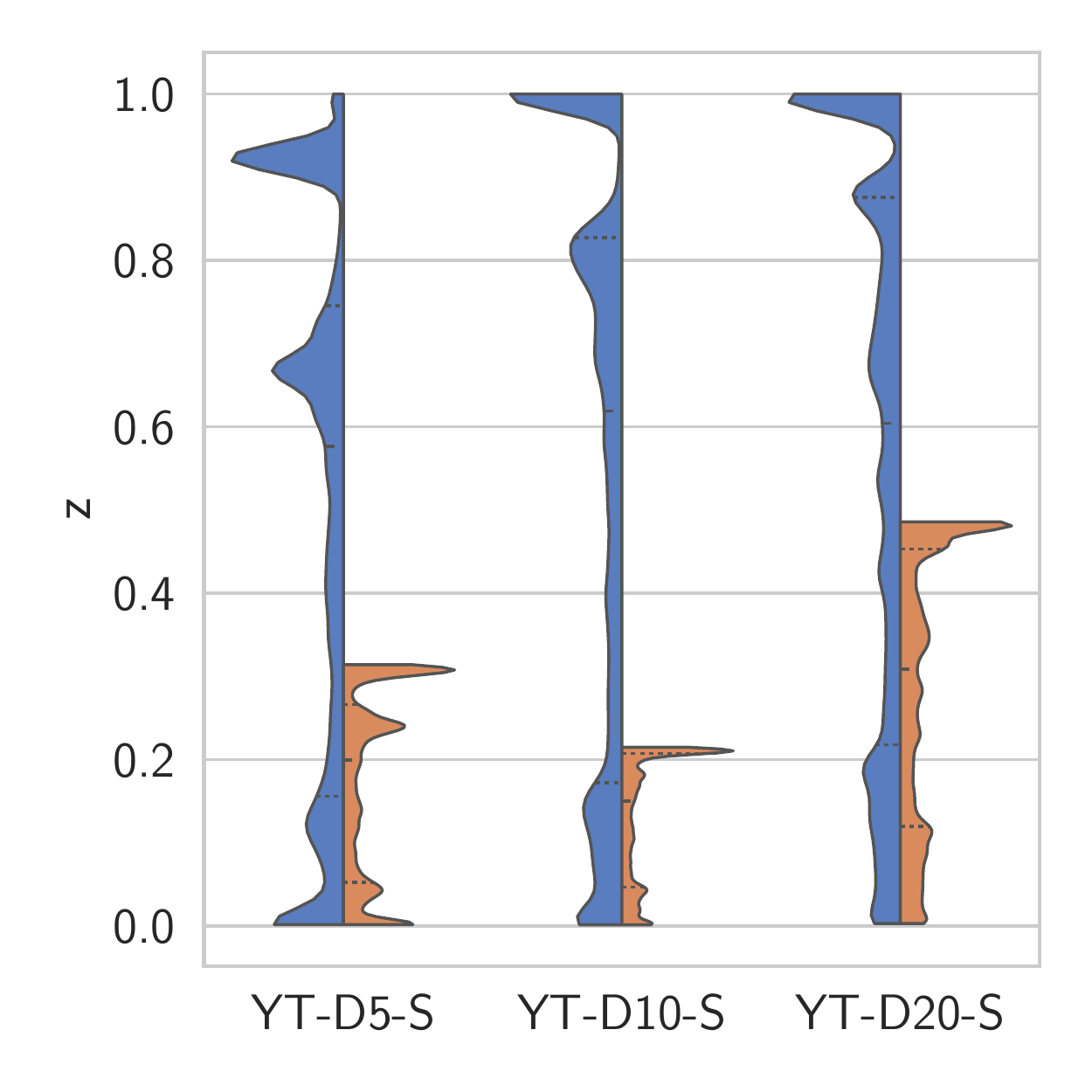}}
  \subcaptionbox{RePBubLik (RBL)}[.475\linewidth]{\includegraphics[width=1.5in]{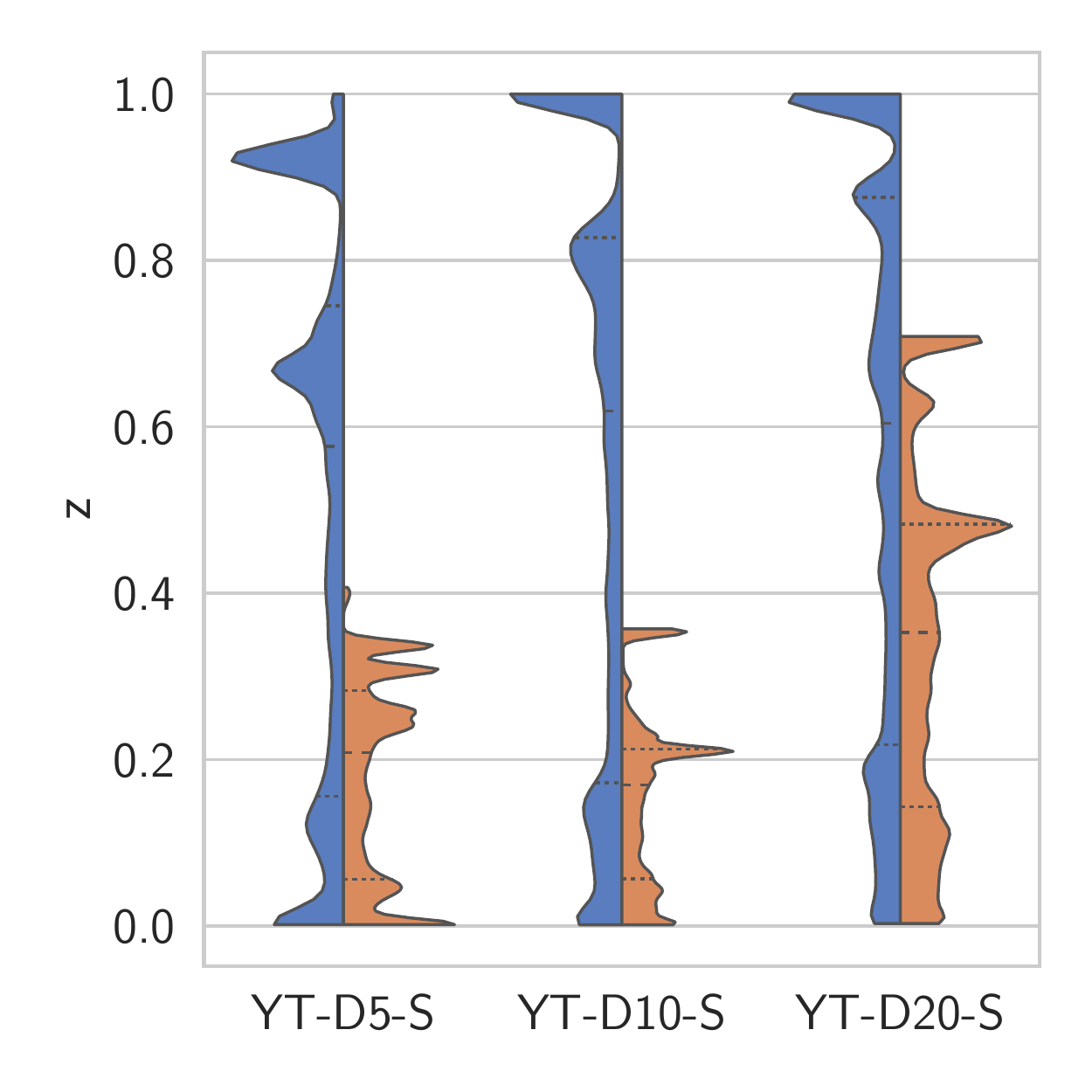}}
  \caption{Distribution of the segregation scores ($z$ values) of harmful nodes before (blue) and after (red) performing 50 rewiring operations provided by HEU and RBL.}
  \Description{Figure 4}
  \label{fig:violinplot}
\end{figure}

\spara{Total exposure to harmful content.}
Having tested the effectiveness of our algorithm in reducing the maximum segregation score, we study its effect on the distribution of the segregation scores over all harmful nodes.
Figure~\ref{fig:violinplot} depicts the distribution of the $z$ values before and after the rewiring operations (with $k = 50$ and $\tau = 0.9$) provided by \textbf{HEU} and \textbf{RBL} on the YouTube recommendation graphs.
For each graph, the violin plot in blue (left) denotes the distribution of segregation scores before the rewiring operations and the one in red (right) the distribution after the rewiring operations.
The range of segregation scores is normalized to $[0,1]$, where the maximum corresponds to the initial segregation.
We observe that reducing the maximum segregation also helps reduce the segregation scores of other harmful nodes. Compared to \textbf{RBL}, \textbf{HEU} generates a distribution more highly concentrated around smaller values; this discrepancy between the distributions is most significant when $d = 20$.

\section{Conclusions and Future Work}
\label{sec:conclusion}

In this paper we studied the problem of reducing the risk of radicalization pathways in  \emph{what-to-watch-next} recommenders via edge rewiring on the recommendation graph.
We formally defined the segregation score of a radicalized node to measure its potential to trap users into radicalization pathways, and formulated the \krewire problem to minimize the maximum segregation score among all radicalized nodes, while maintaining the quality of the recommendations.
We proposed an efficient yet effective greedy algorithm based on the absorbing random walk theory.
Our experiments, in the context of video and news recommendations, confirmed the effectiveness of our proposed algorithm.

This work is just a first step and it has several limitations that we plan to tackle in future work.
One main limitation is assuming a binary labelling of nodes, which limits each content to one of the two groups (harmful or neutral), which is not always realistic.
A natural extension is to assume numerical labels in $[0,1]$. This would require to re-define segregation accordingly.

Another limitation is that we are given the recommendation graph as input. This implies that the recommendations are precomputed and static.
We plan to extend our setting to a scenario where recommendations are generated dynamically in an online fashion.

Finally, we showed through empirical evidence how our method, designed to reduce maximum segregation, may actually reduce the \emph{total} segregation generated by all harmful nodes in the graph. We plan to design different algorithms which are able to directly tackle this objective.

\begin{acks}
Francesco Fabbri is a fellow of Eurecat's \emph{Vicente L\'{o}pez} PhD grant program; his work was partially financially supported by the Catalan Government through the funding grant \emph{ACCI\'{O}-Eurecat} (Project \emph{PRIVany-nom}).
Francesco Bonchi acknowledges support from Intesa Sanpaolo Innovation Center. 
Carlos Castillo has been partially supported by the \emph{HUMAINT} programme (Human Behaviour and Machine Intelligence), European Commission, and by \emph{"la Caixa"} Foundation (ID 100010434), under agreement LCF/PR/PR16/51110009.
Michael Mathioudakis has been supported by the \emph{MLDB} project of the Academy of Finland (decision number: 322046).

The funders had no role in study design, data collection
and analysis, decision to publish, or preparation of the manuscript.
\end{acks}

\balance
\bibliographystyle{ACM-Reference-Format}
\bibliography{biblio}
\clearpage

\balance
\appendix

\section{Proof of Theorem~\ref{thm:np:hard}}
\label{proof:np:hard}

\begin{figure}
  \centering
  \captionsetup[sub]{skip=0pt}
  \subcaptionbox{Construct $G^*$ from $G$\label{subfig:vc:construct}}[.475\linewidth]{\includegraphics[width=1.5in]{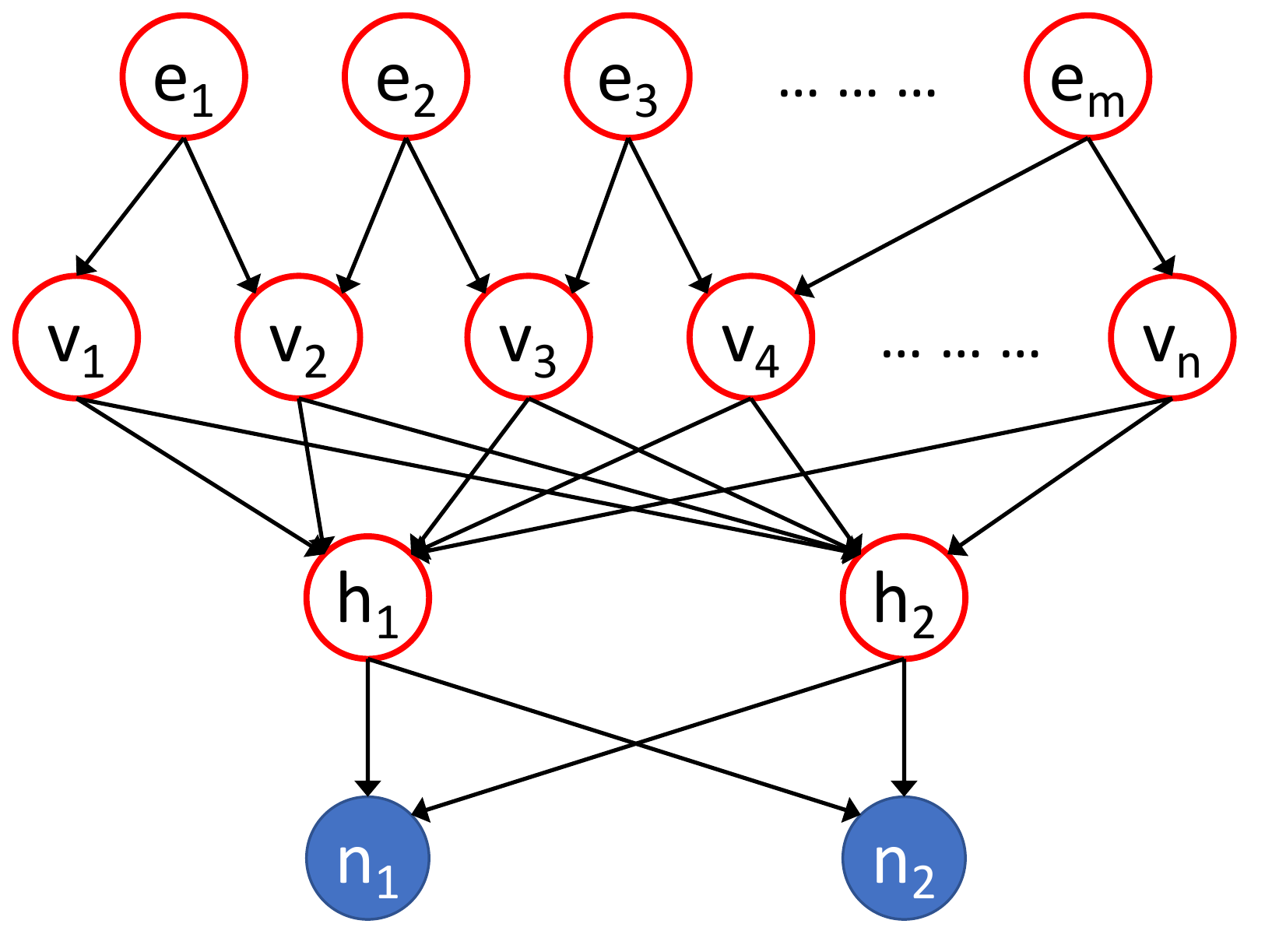}}
  \subcaptionbox{\krewire on $G^*$\label{subfig:vc:rewire}}[.475\linewidth]{\includegraphics[width=1.5in]{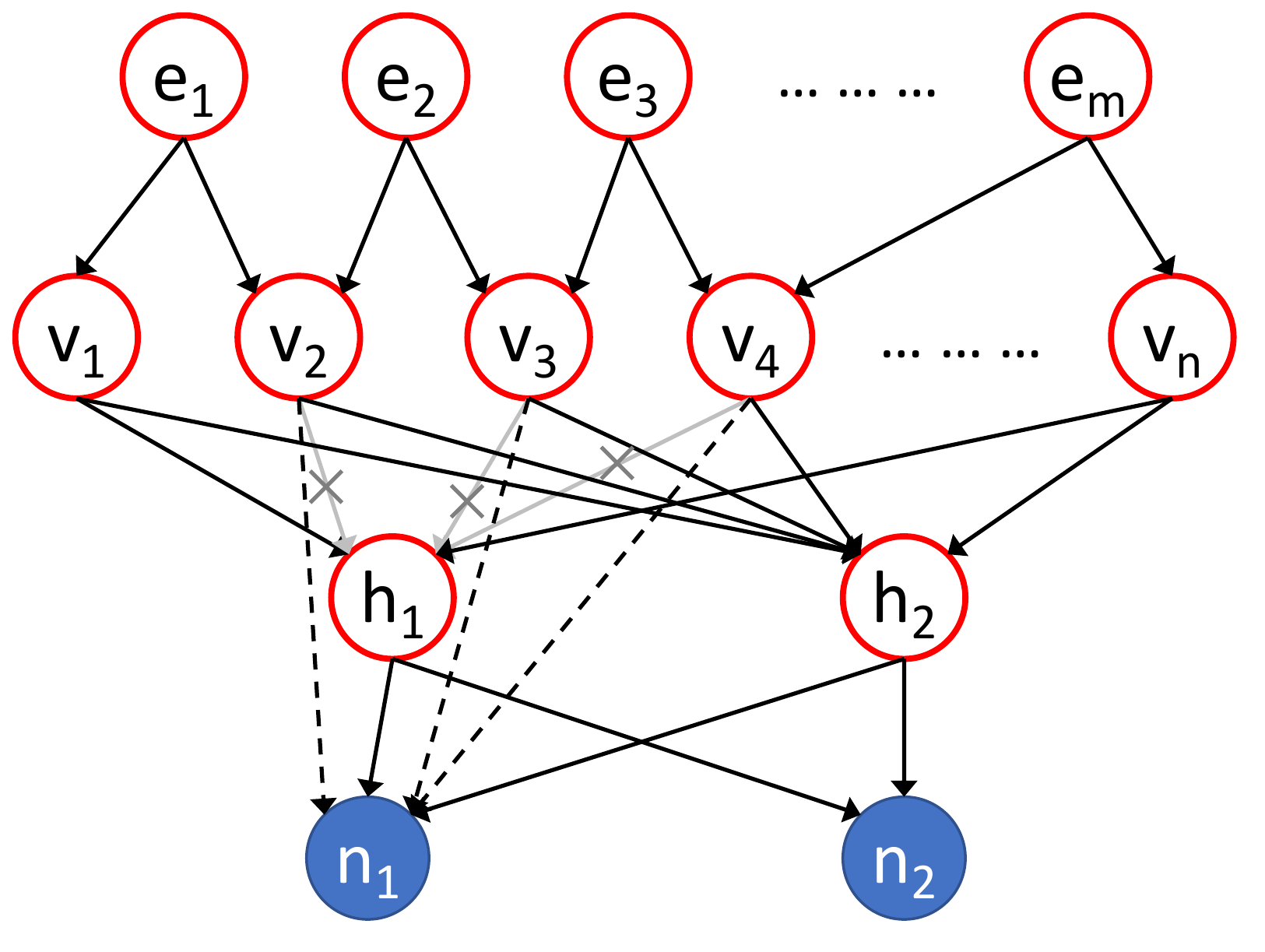}}
  \caption{Illustration of the reduction from the \vertexcover problem to the \krewire problem.}
  \Description{proof of NP-hardness}
  \label{fig:np:proof}
\end{figure}

\begin{proof}
  We prove the NP-hardness of the \krewire problem by a reduction from the \vertexcover problem~\cite{DBLP:books/fm/GareyJ79}.
  
  A \vertexcover instance is specified by an undirected graph $G=(V,E)$, where $|V|=n$ and $|E|=m$, and an integer $k$. It asks whether $G$ has a vertex cover of size at most $k$, i.e., whether there exists a subset $ C \subseteq V $ with $ |C| \leq k $ such that $ \{v_i, v_j\} \cap C \neq \varnothing $ for every edge $e = (v_i, v_j) \in E$. We construct an instance of the \krewire problem on $G^*$ from a \vertexcover instance on $G$ as illustrated in Figure~\ref{subfig:vc:construct}. Given a graph $G=(V,E)$, the graph $G^*=(V^*,E^*)$ is constructed as follows: One vertex in $G^*$ is created for each $e \in E$ and $v \in V$. Furthermore, four vertices $h_1, h_2, n_1, n_2$ are added to $G^*$. Let $V^*_h = E \cup V \cup \{h_1, h_2\}$ be the set of $(m+n+2)$ ``harmful'' vertices (in red) and $V^*_n=\{n_1, n_2\}$ be the set of two ``neutral'' vertices (in blue). Then, for each $e=(v_i, v_j) \in E$, two directed edges $(e, v_i)$ and $(e, v_j)$ are added to $G^*$. For each $v \in V$, two directed edges $(v, h_1)$ and $(v, h_2)$ are added to $G^*$. Finally, four directed edges $(h_1, n_1)$, $(h_1, n_2)$, $(h_2, n_1)$, and $(h_2, n_2)$ are added to $G^*$. The out-degree $d$ of each red node in $G^*$ is $2$. Accordingly, the transition probability of every edge in $G^*$ is set to $0.5$.

  We first show that there will be a set $O$ of at most $k$ rewiring operations such that $\Delta(O) > 0$ after the rewiring operations in $O$ are performed on $G^*$ \emph{if} $G$ has a vertex cover of size at most $k$. For the original $G^*$, we have $z(h_1)=z(h_2)=1$, $z(v)=2$ for each vertex $v \in V$, and $z(e)=3$ for each edge $ e \in E $. Thus, we have $Z=z(e)=3$. So, we will have $\Delta(O) > 0$ as long as $ z'(e) < 3 $ for each edge $e \in E$. Let $C=\{v_1,\ldots,v_k\}$ be a size-$k$ vertex cover of $G$. We construct a set $O=\{o_1,\ldots,o_k\}$ of $k$ rewiring operations on $G^*$, where $ o_i = (v_i, h_1, n_1) $,  corresponding to $C$, as illustrated in Figure~\ref{subfig:vc:rewire}.
  After performing the set $O$ of rewiring operations on $G^*$, we have two cases for $z'(e)$ of each $e=(v_i, v_j)$:
  \begin{equation*}\label{eq:ze}
    z'(e) =
    \begin{cases}
      0.5 \times 3 + 0.5 \times 2 = 2.5, \quad \textit{if} \;\; |\{v_i, v_j\} \cap C| = 2 \\
      0.5 \times 3 + 0.5 \times 2.5 = 2.75, \quad \textit{if} \;\; |\{v_i, v_j\} \cap C| = 1
    \end{cases}
  \end{equation*}
  Since $C$ is a vertex cover, there is no edge $e=(v_i, v_j)$ such that $ \{v_i, v_j\} \cap C = \varnothing $. Therefore, after performing the set $O$ of rewiring operations on $G^*$, it must hold that $z'(e)<3$ for every $e \in E$ and thus $\Delta(O)>0$.

  We then show that there will be a set $O$ of at most $k$ rewiring operations such that $\Delta(O) > 0$ after the rewiring operations in $O$ are performed on $G^*$ \emph{only if} $G$ has a vertex cover of size at most $k$. Or equivalently, if $G$ does not have a vertex cover of size $k$, then any set $O$ of $k$ rewiring operations performed on $G^*$ cannot make $\Delta(O) > 0$. Since $G$ does not have a vertex cover of size $k$, there must exist some edge $ \overline{e}=(v_i, v_j) $ with $ \{v_i, v_j\} \cap \overline{C} = \varnothing $ for any size-$k$ vertex set $\overline{C} \subseteq V$. Therefore, after performing the set $\overline{O}$ of $k$ rewiring operations corresponding to $\overline{C}$, we have $z'(\overline{e}) = 3$ for an uncovered edge $\overline{e}$. So, we can say that any set of $k$ rewiring operations from $V$ cannot make $\Delta(O) > 0$. Furthermore, we consider the case of $k$ rewiring operations from $E$, i.e., to find a set of $k$ edges $\{e_1,\ldots,e_k\}$ and rewire one out-edge from each of them to $n_1$ or $n_2$. In this case, we can always find some unselected edge $\overline{e}$ with $z'(\overline{e})=3$ as long as $ m > k $, which obviously holds as $G$ does not have a vertex cover of size $k$. Finally, we consider the case of a ``hybrid'' set of $k$ rewiring operations from both $E$ and $V$. W.l.o.g., we assume that there are $ (k-k^\prime) $ operations from $V$ and $ k^\prime $ operations from $E$ for some $ 0 < k^\prime < k $. Since $G$ does not have a vertex cover of size $k$, we can say that any vertex set $ \overline{C} $ of size $ (k-k^\prime) $ can cover at most $ (m - k^\prime - 1) $ edges. Otherwise, we would find a vertex cover of size $k$ by adding $ k^\prime $ vertices to cover the remaining $ k^\prime $ edges and thus lead to contradiction. Therefore, after performing only $ k^\prime $ rewiring operations from $E$, there always exists at least one edge $\overline{e}$ that are covered by neither the vertex set nor the edge set, and thus $ z^\prime(\overline{e}) = 3 $ and $\Delta(O) = 0$. Considering all the three cases, we prove that any set of $k$ rewiring operations performed on $G^*$ cannot make $\Delta(O) > 0$ if $G$ does not have a vertex cover of size $k$.

  Given that both the ``\emph{if}'' and ``\emph{only-if}'' directions are proven and $G^*$ can be constructed from $G$ in $O(m+n)$ time, we reduce from the \vertexcover problem to the \krewire problem in polynomial time and thus prove that the \krewire problem is NP-hard.

  To show the hardness of approximation, we suppose that there is a polynomial-time algorithm $\mathcal{A}$ that approximates the \krewire problem within a factor of $\alpha > 0$. Or equivalently, for any \krewire instance, if $ O^* $ is the set of $k$ optimal rewiring operations, then the set $ O^\prime $ of $k$ rewiring operations returned by $\mathcal{A}$ will always satisfy that $\Delta(O^\prime) \geq \alpha \cdot \Delta(O^*)$. Let us consider a \krewire instance on the above graph $G^*$ constructed from $G$ and $k$ be the size of the minimum vertex cover of $G$. For this instance, the optimal solution $O^*$ of the \krewire problem exactly corresponds to the minimum vertex cover $C^*$ of $G$ with $ \Delta(O^*) > 0 $; any other solution $ O^\prime $ will lead to $ \Delta(O^\prime) = 0 $, as we have shown in this proof. If $ \mathcal{A} $ could find a solution for the \krewire problem with any approximation factor $\alpha > 0$ in polynomial time, then $\mathcal{A}$ would have solved the \vertexcover problem in polynomial time, which has been known to be impossible unless P=NP. Therefore, the \krewire problem is NP-hard to approximate with any factor.
\end{proof}

\section{Ethics Statement}

In this work, we aim at reducing the exposure to radicalized content generated by W2W recommender systems. 
Our approach does not include any form of censorship, and instead limits algorithmic-induced over-exposure, which is stimulated by biased organic interactions (e.g., the spread of radicalized content through user-user interactions).
Our work contributes to raise awareness on the importance of devising policies aimed at reducing harmful algorithmic side-effects.
Generally, we do not foresee any immediate and direct harmful impacts from this work.

\end{document}